\documentclass[floats,prb,twocolumn]{revtex4}
\usepackage{epsfig}
\usepackage{graphicx}
\usepackage{color}

\def\be{\begin{equation}}
\def\ee{\end{equation}}
\def\ba#1{\begin{array}{#1}}
\def\ea{\end{array}}
\def\bn{\begin{enumerate}}
\def\en{\end{enumerate}}

\def\rr{\right}
\def\l{\left}

\def\ket#1{\l|#1\rr\rangle}
\def\bra#1{\l\langle#1\rr|}

\def\beq{\begin{equation}}
\def\eeq{\end{equation}}

\begin{document}

\title{Many-Body Localization in a Quasiperiodic System}
\author{Shankar Iyer}
\author{Gil Refael}
\affiliation{Department of Physics, California Institute of
  Technology, MC 149-33, 1200 E.\ California Blvd., Pasadena, CA 91125}
\author{Vadim Oganesyan}
\affiliation{Department of Engineering Science and Physics,
College of Staten Island, CUNY, Staten Island, NY 10314}
\affiliation{
The Graduate Center, CUNY, 365 5th Ave., New York, NY, 10016}
\affiliation{
KITP, UCSB, Santa Barbara, CA 93106-4030}
\author{David A. Huse}
\affiliation{Department of Physics, Princeton University, Princeton, NJ 08544}

\date{\today}

\begin{abstract}
Recent theoretical and numerical evidence suggests that localization can survive in disordered many-body systems with very high energy density, provided that interactions are sufficiently weak.  Stronger interactions can destroy localization, leading to a so-called many-body localization transition.  This dynamical phase transition is  relevant to questions of thermalization in extended quantum systems far from the zero-temperature limit.  It separates a many-body localized phase, in which localization prevents transport and thermalization, from a conducting (``ergodic'') phase in which the usual assumptions of quantum statistical mechanics hold.  Here, we present numerical evidence that many-body localization also occurs in models without disorder but rather a quasiperiodic potential.  In one dimension, these systems already have a single-particle localization transition, and we show that this transition becomes a many-body localization transition upon the introduction of interactions.  
We also comment on possible relevance of our results to experimental studies of many-body dynamics of cold atoms and non-linear light in quasiperiodic potentials.
\end{abstract}

\maketitle

\section{Introduction}

\indent In one-dimensional systems of non-interacting particles, an arbitrarily weak disordered potential generically localizes all quantum eigenstates~\cite{anderson1958absence,abrahams1979scaling}.  Such a system is always an insulator, with a vanishing conductivity in the thermodynamic limit.  The question of how this picture is modified by interactions remained unclear in the decades following Anderson's original work on localization~\cite{fleishman1980interactions,thouless1981conductivity}.  Relatively recently, Basko, Aleiner, and Altshuler have argued that an interacting many-body system can undergo a so-called many-body localization (MBL) transition in the presence of quenched disorder.  At low energy density and/or strong disorder, interactions are insufficient to thermalize the system, so the system remains a ``perfect" insulator (i.e. with zero DC conductivity despited being excited); at higher energy density and/or weaker disorder, the conductivity can become nonzero and the system thermalizes, leading to a conducting phase~\cite{basko2006metal,basko2007problem}.

\indent The MBL transition is rather unique for several reasons.  First, in contrast to more conventional quantum phase transitions\cite{sachdev2007quantum}, this is not a transition in the ground state.  Instead, the MBL transition involves the localization of highly excited states of a many-body system, with finite energy density.  This means that the transition differs from most metal-insulator transitions, which are sharp only at zero temperature\cite{dobrosavljevic2011introduction}.  Furthermore, this MBL transition is of fundamental interest in the context of statistical mechanics.  Local subsystems of interacting, many-body systems are generically expected to equilibrate with their surroundings, with statistical properties of these subsystems reaching thermal values after sufficient time.  Studies of how this occurs in quantum systems have led to the so-called \textit{eigenstate thermalization hypothesis} (ETH), which states that individual eigenstates of the interacting quantum system already encode thermal distributions of local quantities~\cite{deutsch1991quantum,srednicki1994chaos}.  However, the many-body localized phase provides an example of a situation in which the ETH is false, and the ergodic hypothesis of quantum statistical mechanics is violated~\cite{oganesyan2007localization,pal2010many}.  Since the work of Basko et al., these intriguing aspects of MBL have motivated many studies aimed at locating and understanding this transition in disordered systems~\cite{oganesyan2007localization,znidaric2008many,karahalios2009finite,oganesyan2007energy,pal2010many,monthus2010many,berkelbach2010conductivity,canovi2011quantum,bardarson2012unbounded,
vosk2012many,de2012many,khatami2012quantum,biroli2012difference}.

\indent On the other hand, it is important to note that single-particle localization does not require disorder.  In 1980, Aubry and Andr\'{e} studied a 1D single-particle tight-binding model that omits disorder in favor of a potential that is periodic, but with a period that is incommensurate with the underlying lattice\cite{aubry1980analyticity}.  Harper had studied a similar model much earlier, but he had focused on a special ratio of hopping to potential strength\cite{harper1955single}.  Aubry and Andr\'{e} showed that this point actually lies at a localization transition.  It separates a weak potential phase, where all single-particle eigenstates are extended, from a strong potential phase, where all eigenstates are localized.  In the 1980s and 1990s, physicists continued to study this quasiperiodic localization transition for its own peculiarities and because it mimics the situation in disordered systems in $d \geq 3$, where there is also a single-particle localization transition~\cite{thouless1983wavefunction,chaves1997transport,evangelou1997two,abanov1998asymptotically,eilmes1999two,siebesma1987multifractal,hashimoto1992finite,vidal1999correlated}.  The AA model was also actively investigated in the mathematical physics literature, because it involves a Schr\"{o}dinger operator (i.e. the ``almost Mathieu" operator) with particularly rich spectral properties.  The contributions of mathematical physicists put the initial work on Aubry and Andr\'{e} on more rigorous footing~\cite{simon1982almost,bellissard1983metal,jitomirskaya1994operators,jitomirskaya1999metal}.  More recently, the AA model has been directly experimentally realized in cold atom experiments\cite{fallani2007ultracold,lucioni2011observation} and also in photonic waveguides\cite{lahini2009observation}.  The possibility of engineering quasiperiodic systems in the laboratory has inspired new theoretical and numerical work aimed at understanding the localization properties of such systems and how they differ from those with true disorder~\cite{albert2010localization,cestari2011critical, nessi2011finite,tezuka2012testing,he2012noise,ribeiro2012strongly,gramsch2012dynamics}.

\subsubsection{Statement of the problem and summary of the results}

\indent In this paper, we ask whether there can be a MBL transition in an interacting extension of the AA model.
More concretely, suppose we begin with a half-filled, one-dimensional system of fermions or hardcore bosons in a particular randomly chosen many-body Fock state, with some sites occupied and others empty.
Such a configuration of particles is typically far from the ground state of the system.  Instead, by sampling the initial configuration uniformly at random (i.e. without regard to its energy content), 
we are actually working in the so-called \textit{infinite temperature limit}.  If the particles are allowed to hop and interact for a sufficiently long time, the standard expectation is that 
the system should \textit{thermalize}: that is, all microscopic states that are consistent with conservation laws should become equally likely and local observables should thereby assume some thermal distribution\cite{rigol2008thermalization}.
Can this expectation be violated in the presence of a quasiperiodic potential?  In other words, can the system fail to serve as a good heat bath for itself?  If so, can this be traced to the persistence of localization even in the presence of interactions?

\begin{figure}
\includegraphics[width=7.8cm]{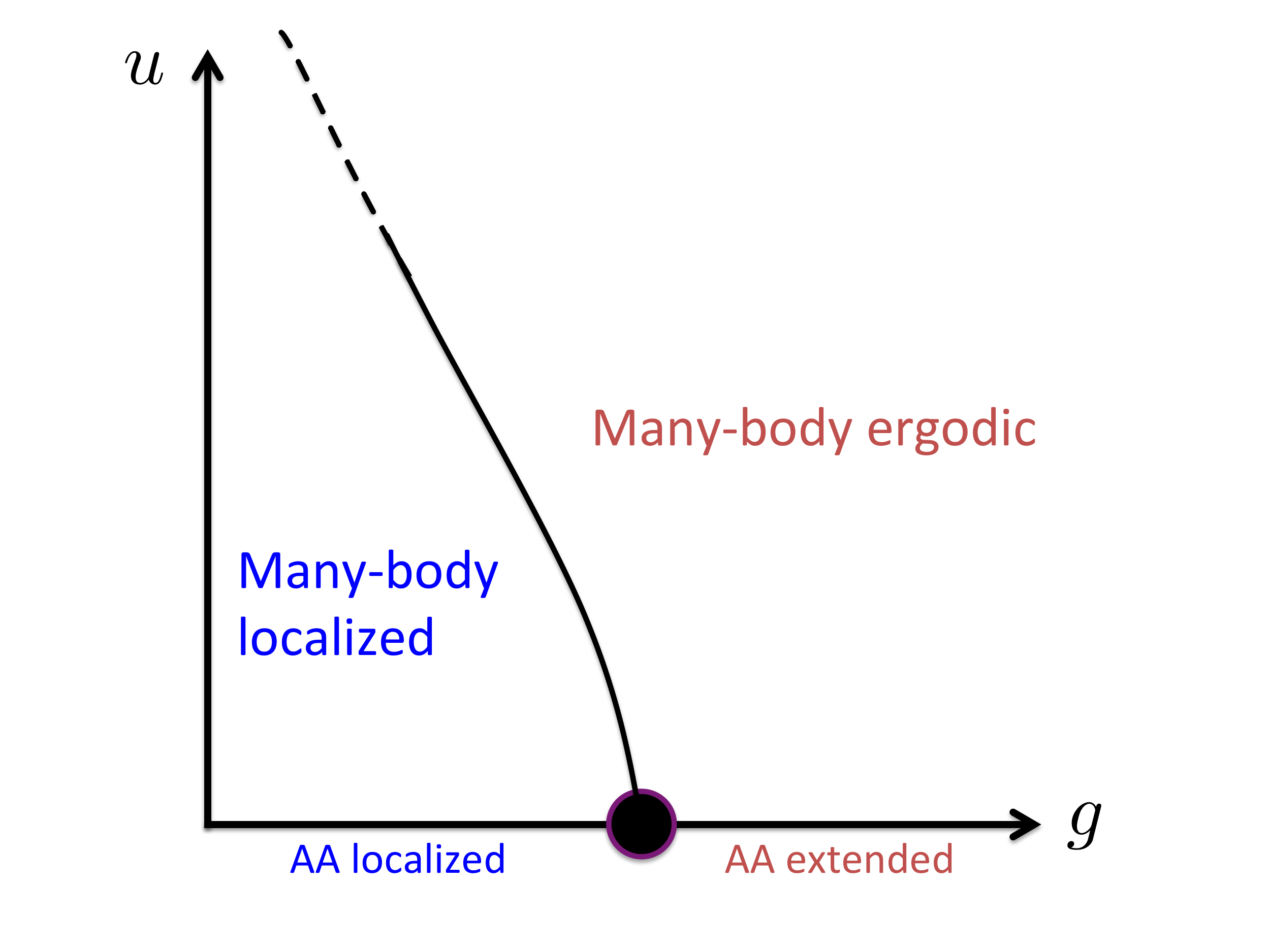}
\caption{The proposed phase diagram of our interacting Aubry-Andr\'{e} model at high energy density.  Interactions convert the localized and extended phases of the AA model into many-body localized and ergodic phases and induce an
expansion of the many-body ergodic phase.  The phases of the interacting model differ qualitatively from their non-interacting counterparts.  The differences are explained in Section \ref{sec:discussion} below.  }
\label{fig:phasediagram}
\end{figure}

\indent The answer to  both of these questions appears to be ``yes."  We use numerical simulations of unitary evolution
of a many-body quasiperiodic system to measure three kinds of observables in the limit of very late times: the correlation between the initial and time-evolved particle density profiles, the many-body participation ratio, and the R\'{e}nyi entropy.  
Our observations are consistent with the existence of two phases in the parameter space of our model that differ qualitatively in \textit{ergodicity}.  
At finite interparticle interaction strength $u$ and large hopping $g$,
there exists a phase in which the usual assumptions of statistical mechanics appear to hold.  The initial state evolves into a superposition of a finite fraction of the
total number of possible configurations, and consequently, local observables approximately assume their thermal values.  This is the many-body ergodic phase.
However, at small hopping $g$, there is a phase in which particle transport away from the initial configuration is not strongly enhanced by interactions.  The system
explores only an exponentially small fraction of configuration space, and local observables do not even approximately thermalize.  This is the many-body localized phase.
\indent Figure \ref{fig:phasediagram} presents a schematic illustration of the proposed phase diagram.  Although interactions induce an expansion of the ergodic regime, the localized phase
survives at finite $u$, and consequently, there is evidence for a quasiperiodic MBL transition\footnote{Both the many-body ergodic and localized phases differ \emph{qualitatively} from their counterparts in the \emph{non-interacting} AA model.  The non-interacting extended phase is not ergodic, indicating
that interactions are necessary for thermalization.  Meanwhile, the many-body localized phase is expected to exhibit logarithmic growth of the bipartite entanglement entropy to an extensive value, albeit with subthermal entropy density.  Such behavior is in fact consistent with the recent observations in the disordered problem\cite{znidaric2008many,bardarson2012unbounded}.  This growth is absent in the AA localized phase without interactions.  Despite this difference, the interacting and non-interacting localized phases
are similar in their inability to thermalize the particle density.}.

\indent There has certainly been substantial previous work on localization in many-body quasiperiodic systems.  For instance, Vidal et al.\cite{vidal1999correlated} adapted the approach of Giamarchi and Schulz\cite{giamarchi1988anderson} to study the effects of a perturbative quasiperiodic potential on the low-energy physics of interacting fermions in one dimension.  Very recently, He et al.\cite{he2012noise} studied the ground state Bose glass to superfluid transition for hardcore bosons in a 1D quasiperiodic lattice.  Our work differs fundamentally from these and many other studies precisely because it focuses on non-equilibrium behavior in the high-energy (infinite temperature) limit and argues that a localization transition can even occur in this regime.

\subsubsection{Organization of the paper}

\indent We begin our study in Section \ref{sec:model} by introducing our interacting extension of the standard AA model.  Since the MBL transition is a non-equilibrium phase transition, our goal is to follow the real-time dynamics.  To simplify this task, we describe a method of modifying the dynamics of our model, such that numerical integration of the new dynamics is somewhat easier than the original problem.  In Section \ref{sec:numerics}, we introduce the quantities that we measure in our simulations and present the numerical results.  Then, in Section \ref{sec:discussion}, we argue that our data points to the existence of many-body localized and many-body ergodic phases by proposing model late-time states for each of these regimes and comparing to the numerical results from Section \ref{sec:numerics}.  Next, in Section \ref{sec:boundary}, we extract estimates for the phase boundary from our data, motivating the phase diagram in Figure \ref{fig:phasediagram}.  Finally, we conclude in Section \ref{sec:conclusion} by summarizing our results, drawing connections to theory and experiment, and suggesting avenues for future extensions of our work.  

\indent We relegate two exact diagonalization studies to Appendix \ref{sec:floquet}.  First, we examine the impact of our modified dynamics upon the single-particle and many-body problems.  Second, we study the many-body level statistics of the interacting model.  We find evidence for a crossover between Poisson and Wigner-Dyson statistics, consistent with the usual expectation in the presence of a localization transition~\cite{shklovskii1993statistics}. 

\section{Model and Methodology}
\label{sec:model}

\indent In this section, we motivate and introduce our model and our numerical methodology for studying real-time dynamics.

\subsection{The ``Parent" Model}

\indent We would like to consider one-dimensional lattice models of the following general form:
\begin{equation}
\label{eq:parentmodel}
\hat{H} = \sum_{j=0}^{L-1} \left[ h_j \hat{n}_j + J(\hat{c}^\dagger_j \hat{c}_{j+1}+\hat{c}^\dagger_{j+1} \hat{c}_j) + V \hat{n}_j \hat{n}_{j+1}\right]
\end{equation}
Here, $\hat{c}_j$ is a fermion annihilation operator, and $\hat{n}_j \equiv \hat{c}^\dagger_{j} \hat{c}_j$ is the corresponding fermion number operator.  The three terms in the Hamiltonian (\ref{eq:parentmodel})
then correspond to an on-site potential, nearest-neighbor hopping, and nearest-neighbor interaction respectively.  For now, we leave the boundary conditions unspecified.   In 1D, the Hamiltonians (\ref{eq:parentmodel}) 
for hardcore bosons and fermions differ only in the matrix elements describing hopping over the boundary.  With open boundary conditions, the Hamiltonians (and consequently all properties of the spectra) are identical.

\indent If we set $V = 0$ in the Hamiltonian (\ref{eq:parentmodel}) and take $h_j$ to be genuinely disordered, we recover the non-interacting Anderson Hamiltonian.  If we then turn on a finite $V = J$, we obtain a model that is related to the spin models that have been studied in the context of MBL\cite{pal2010many,bardarson2012unbounded}.  Alternatively, suppose we set $V = 0$ again and take:
\begin{equation}
\label{eq:aapotential}
h_j = h\cos(2\pi k j + \delta)
\end{equation}
With a generic irrational wavenumber $k$ and an arbitrary offset $\delta$, we obtain the non-interacting AA model\cite{aubry1980analyticity}.  For our purposes, we would like to use an incommensurate potential of the form (\ref{eq:aapotential}), with $h = 1$ and $g \equiv \frac{J}{h}$ and $u \equiv \frac{V}{h}$ left as tuning parameters to explore different phases of the model (\ref{eq:parentmodel}).

\indent Before proceeding, we should briefly review what is known about the single-particle AA model.  With periodic boundary conditions and $\delta = 0$, this model is self-dual~\cite{aubry1980analyticity,albert2010localization}.  The self-duality can be realized by switching to Fourier space ($c_j = \frac{1}{\sqrt{L}}\sum_q e^{iqj} c_q$) and then performing a rearrangement of the wavenumbers $q$ such that the real-space potential term looks like a nearest-neighbor hopping in Fourier space and vice versa.  On a finite lattice of length $L$ with periodic boundary conditions, such a rearrangement is possible whenever the wavenumber of the \textit{potential} $k = \frac{\ell}{L}$ such that $\ell$ and $L$ are mutually prime.  The duality construction reveals that, if the AA model has a transition, it must occur at $g = \frac{1}{2}$.  In the thermodynamic limit, there is indeed a transition at this value for nearly all irrational wavenumbers $k$\cite{thouless1983wavefunction}.  When $g > \frac{1}{2}$, all single-particle eigenstates are spatially extended, and by duality, localized in momentum space; when $g < \frac{1}{2}$, all single-particle eigenstates are spatially localized, and by duality, extended in momentum space.  Exactly at $g = \frac{1}{2} $, the eigenstates are multifractal~\cite{siebesma1987multifractal,hashimoto1992finite}.  The spatially extended phase of the AA model is characterized by ballistic, not diffusive, transport \cite{aubry1980analyticity}.  Recently, Albert and Leboeuf have argued that localization in the AA model is a fundamentally more classical phenomenon than disorder-induced Anderson localization, and that the AA transition at $g = \frac{1}{2}$ is most simply viewed as the classical trapping that occurs when the maximum eigenvalue of the kinetic (or hopping) term crosses the amplitude of the incommensurate potential\cite{albert2010localization}.

\subsection{Numerical Methodology and Modification of the Quantum Dynamics}

\indent Probing the MBL transition necessarily involves studying highly excited states of the system, and this precludes the application of much of the extensive machinery that has been developed for investigating low-energy physics.  Consequently, several studies of MBL have resorted to exact diagonalization or other methods involving similar numerical cost~\cite{oganesyan2007localization,pal2010many,monthus2010many}.  We too use a numerical methodology that scales exponentially in the size of the system.  However, in order to access longer evolution times in larger lattices, we introduce a modification of the quantum dynamics.  This modification is inspired by a scheme used previously by 
two of us in a study of classical spin chains\cite{oganesyan2007energy}.  There, at any given time, either the even spins in the chain were allowed to evolve under the influence of the odd spins or vice versa.  This provided access to late times that would have been too difficult to access by direct integration of the standard classical equations of motion.  

\indent By analogy, we propose allowing hopping on \textit{each bond} in turn.  At any given time, the instantaneous Hamiltonian looks like:
\begin{equation}
\label{eq:singlebondH}
\hat{H}_m =  La_mJ(\hat{c}^\dagger_m \hat{c}_{m+1}+\hat{c}^\dagger_{m+1} \hat{c}_m) + \sum_{j=0}^{L-1} \left[ h_j \hat{n}_j + V \hat{n}_j \hat{n}_{j+1}\right]
\end{equation}
We will specify the value of $a_m$ in Section \ref{sec:model}.C below, where we discuss our choice of boundary conditions.
The state of the system is allowed to evolve under this Hamiltonian for a time $\frac{\Delta t}{L}$, and this evolution can be implemented by applying the unitary operator:
\begin{equation}
\label{eq:singlebondU}
\hat{U}_m =  \exp{\left(-i\frac{\Delta t}{L}\hat{H}_m\right)}
\end{equation}
One full time-step of duration $\Delta t$ consists of cycling through all the bonds:
\begin{equation}
\label{eq:onetimestep}
\hat{U}(\Delta t) =  \prod_{m = 0}^{L-1} \hat{U}_m
\end{equation}
Note that, in (\ref{eq:singlebondH}), the hopping is enhanced by $L$ because the hopping on any given bond is activated only once per cycle, while the potential and interaction terms always act.  Therefore, the factor of $L$ ensures that the average Hamiltonian over a time $\Delta t$ has the form (\ref{eq:parentmodel}).  The advantage of employing the modified dynamics is that the $\hat{H}_m$ only couple pairs of configurations, so preparing the $\hat{U}_m$ reduces to exponentiating order $V_H$ two-by-two matrices, where $V_H$ is the size of the Hilbert space.  This is generally a simpler task than exponentiating the original Hamiltonian (\ref{eq:parentmodel}).  Our scheme only constitutes a polynomial speedup over exact diagonalization, but that speedup can increase the range of accessible lattice sizes by a few sites.

\indent The modified dynamics raise several important issues that should be discussed\cite{maricq1982application}.  The periodic time-dependence of the Hamiltonian induces so-called ``multi-photon" (or ``energy umklapp") transitions between states of the ``parent" model (\ref{eq:parentmodel}) that differ in energy by $\omega_H = \frac{2\pi}{\Delta t}$, reducing energy conservation to \textit{quasienergy} conservation modulo $\omega_H$.  We need to question whether this destroys the physics of interest: does the single-particle Aubry-Andr\'{e} transition survive, or do the multi-photon processes destroy the localized phase?  

\indent We take up this question in Appendix \ref{sec:floquet}, where we present a Floquet analysis of the single-particle and many-body problems.  We find that, for sufficiently small $\Delta t$, the universal behavior of the single-particle AA model is preserved.  At larger $\Delta t$, multi-photon processes can strongly mix eigenstates of the single-particle parent model, increasing the single-particle density-of-states and destroying the AA transition.  In the spirit of the earlier referenced work on classical spin chains\cite{oganesyan2007energy}, our perspective in this paper is to identify whether MBL can occur in a model \textit{qualitatively} similar to our parent model (\ref{eq:parentmodel}).  Therefore, to explore dynamics on long time scales, we avoid destroying the single-particle transition, but still choose $\Delta t$ to be quite large within that constraint.   

\indent In Appendix \ref{sec:floquet}, we also examine the consequences of our choice of $\Delta t$ for the quasienergy spectrum of the many-body model.  Our results suggest that  multi-photon processes do not, in fact, strongly modify the parent model's spectrum for much of the parameter range that we explore in this paper\footnote{There is an exception to this statement: multi-photon processes do seem to play an important role deep in the ergodic phase, where the energy content of the system is especially high.  See Appendix \ref{sec:floquet} and the discussion of the time-dependence of the autocorrelator $\chi$ in Section \ref{sec:numerics}.A for more details.}.  This means that partial energy conservation persists in our simulations despite the introduction of a time-dependent Hamiltonian, and we need to keep this fact in mind when we analyze our numerical data below.   

\indent Finally, we note in passing that several recent studies have focused on the localization properties of time-dependent models\cite{kitagawa2012photo,martinez2006localization,dalessio2012many}, including one on the quasiperiodic Harper model\cite{kolovsky2012driven}, but that the intricate details of this topic are somewhat peripheral to our main focus.

\subsection{Details of the Numerical Calculations}

\indent In studies of the 1D AA model, it is conventional to approach the thermodynamic limit by choosing lattice sizes according to the Fibonacci series ($L = \ldots 5$, $8$, $13$, $21$, $34\ldots$) and wavenumbers for the potential (\ref{eq:aapotential}) as ratios of successive terms in the series\cite{thouless1983wavefunction}.  These values of $k$ respect periodic boundary conditions while converging to the inverse of the golden ratio $\frac{1}{\phi} = 0.618033\ldots$.  For any finite lattice, the potential is only commensurate with the entire lattice (since successive terms in the Fibonacci series are mutually prime), and the duality mapping of the AA model is always exactly preserved.  For our purposes however, this approach offers too few accessible system sizes and complicates matters by generating odd values of $L$.  

\indent Instead, we found empirically that finite-size effects are least problematic if we use exclusively even $L$, always keep the wavenumber of the potential fixed at $k = \frac{1}{\phi}$, and set:
\begin{equation}
 a_m = 1-\delta_{m,L-1}
\label{eq:ambc}
\end{equation} 
in equation (\ref{eq:singlebondH}), thereby forbidding hopping over the boundary\footnote{To appropriately realize open boundary conditions, we should also turn off interactions over the boundary.  When exploring different options for the boundary conditions, we varied $J$ over the boundary and neglected to vary $V$.  This is unfortunate in that it makes the model somewhat stranger.  However, our boundary conditions are chosen for convenience anyway, and the numerics suggest that the choice of boundary conditions does not impact the essential physics discussed in this paper.}.  Note that, with these boundary conditions, our model describes hardcore bosons as well as fermions.  The bosonic language maintains closer contact with cold atom experiments\cite{fallani2007ultracold}; the fermionic language is more in keeping with the MBL literature~\cite{basko2006metal,oganesyan2007localization}.

\indent Using the approach described above, we have simulated systems up to size $L = 20$ at half-filling.  Our simulations always begin with a randomly chosen configuration (or Fock) state, so that the initial state has no entanglement across any spatial bond in the lattice (i.e. each site is occupied or empty with probability 1).  Except in the exact diagonalization studies of Appendix \ref{sec:floquet}, we always set $\Delta t = 1$.  We integrate out to $t_f = 9999$ and ultimately average the evolution of measurable quantities over several \textit{samples}, where a sample is specified by the choice of the initial configuration and offset phase to the potential (\ref{eq:aapotential}).   The sample counts used in the numerics are provided in Table \ref{tab:samples}.

\begin{table}
\centering
\begin{tabular}{|r|r|r|r|r|}
\hline
$L$ & $N$ & $V_{H}$ & samples\\
\hline
$8$ & $4$ & $70$ & $500$  \\
\hline
$10$ & $5$ & $252$ & $500$  \\
\hline
$12$ & $6$ & $924$ & $500$  \\
\hline
$14$ & $7$ & $3432$ & $250$  \\
\hline
$16$ & $8$ & $12870$ & $250$  \\
\hline
$18$ & $9$ & $48620$ & $250$  \\
\hline
$20$ & $10$ & $184756$ & $50$  \\
\hline
\end{tabular}
\caption{For the various simulated lattice sizes $L$, the particle number $N$, the configuration space size $V_H$, and the number of samples used in the numerics.  Note that we always work at half-filling.}
\label{tab:samples}
\end{table}

\section{Numerical Measurements}
\label{sec:numerics}

\indent We now introduce the quantities that we measure to characterize the different regimes of our model.  We also present the numerical data along with some qualitative
remarks about the observed behavior.  However, we largely defer quantitative phenomenology and modeling of the data to Section \ref{sec:discussion}.

\subsection{Temporal Autocorrelation Function}

\indent One signature of localization is the system's retention of memory of its initial state.  Since we simulate the reversible evolution of a closed system, the quantum state of the entire system retains full memory of its past.  Nevertheless, we may still ask if the information needed to deduce the initial state is preserved locally or if it propagates to distant parts of the system.  A diagnostic measure with which to pose this ``local memory" question is the \textit{temporal autocorrelator} of site $j$:
\begin{equation}
\label{eq:temporalacsite}
\chi_j(t) \equiv (2\langle \hat{n}_j \rangle(t) -1)(2\langle \hat{n}_j \rangle (0) -1)
\end{equation}
Here, the angular brackets refer to an expectation value in the quantum state.  This single-site autocorrelator may be averaged over sites and then over samples (as defined in Section \ref{sec:model}.C) to obtain:
\begin{equation}
\label{eq:temporalac}
\chi(t;L) \equiv \left [ \frac{1}{L} \sum_{j = 0}^{L-1} \chi_j(t) \right ]
\end{equation}
The sample average is indicated here with the large square brackets.
Typically, to reduce the effects of noise, we also average over a few time steps within each sample (i.e.\ perform \textit{time binning}) before taking the sample average.

\begin{figure}
\begin{minipage}[b]{0.4cm}
       {\bf (a)}

       \vspace{3.3cm}
\end{minipage}
\begin{minipage}[t]{7.9cm}
       \includegraphics[width=7.8cm]{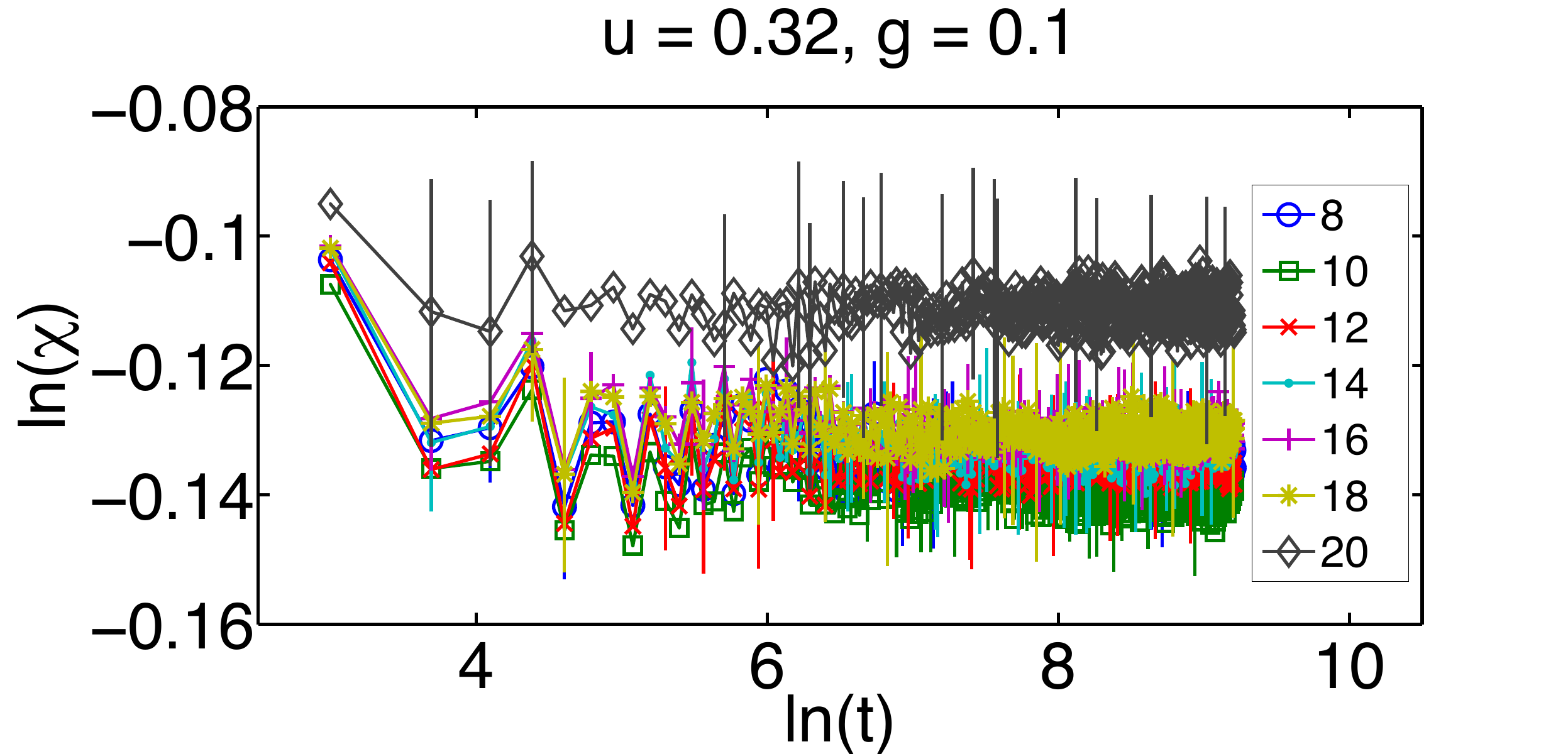}
\end{minipage}\\
\begin{minipage}[b]{0.4cm}
       {\bf (b)}

       \vspace{3.3cm}
\end{minipage}
\begin{minipage}[t]{7.9cm}
       \includegraphics[width=7.8cm]{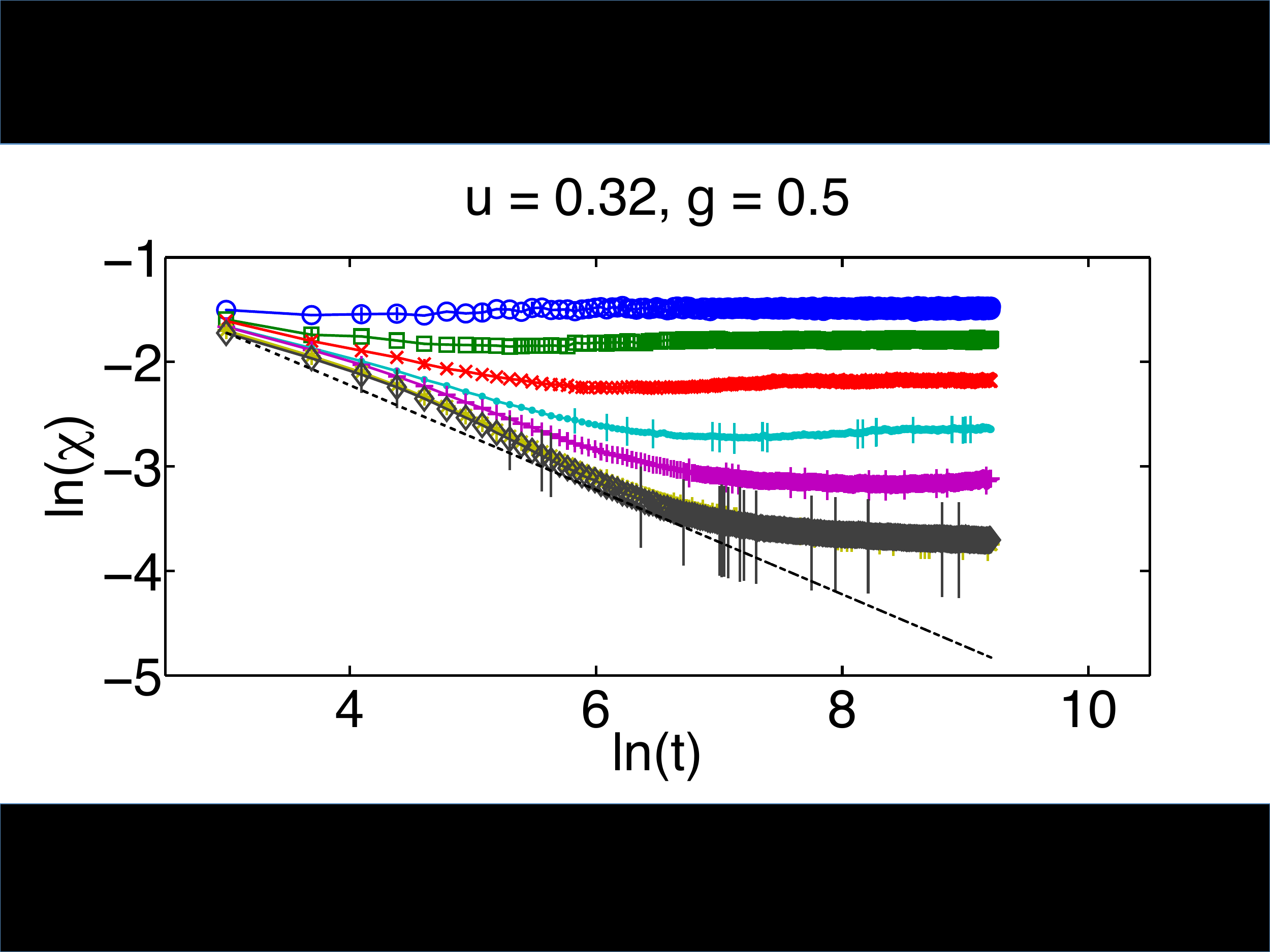}
\end{minipage}\\
\begin{minipage}[b]{0.4cm}
       {\bf (c)}

       \vspace{3.3cm}
\end{minipage}
\begin{minipage}[t]{7.9cm}
       \includegraphics[width=7.8cm]{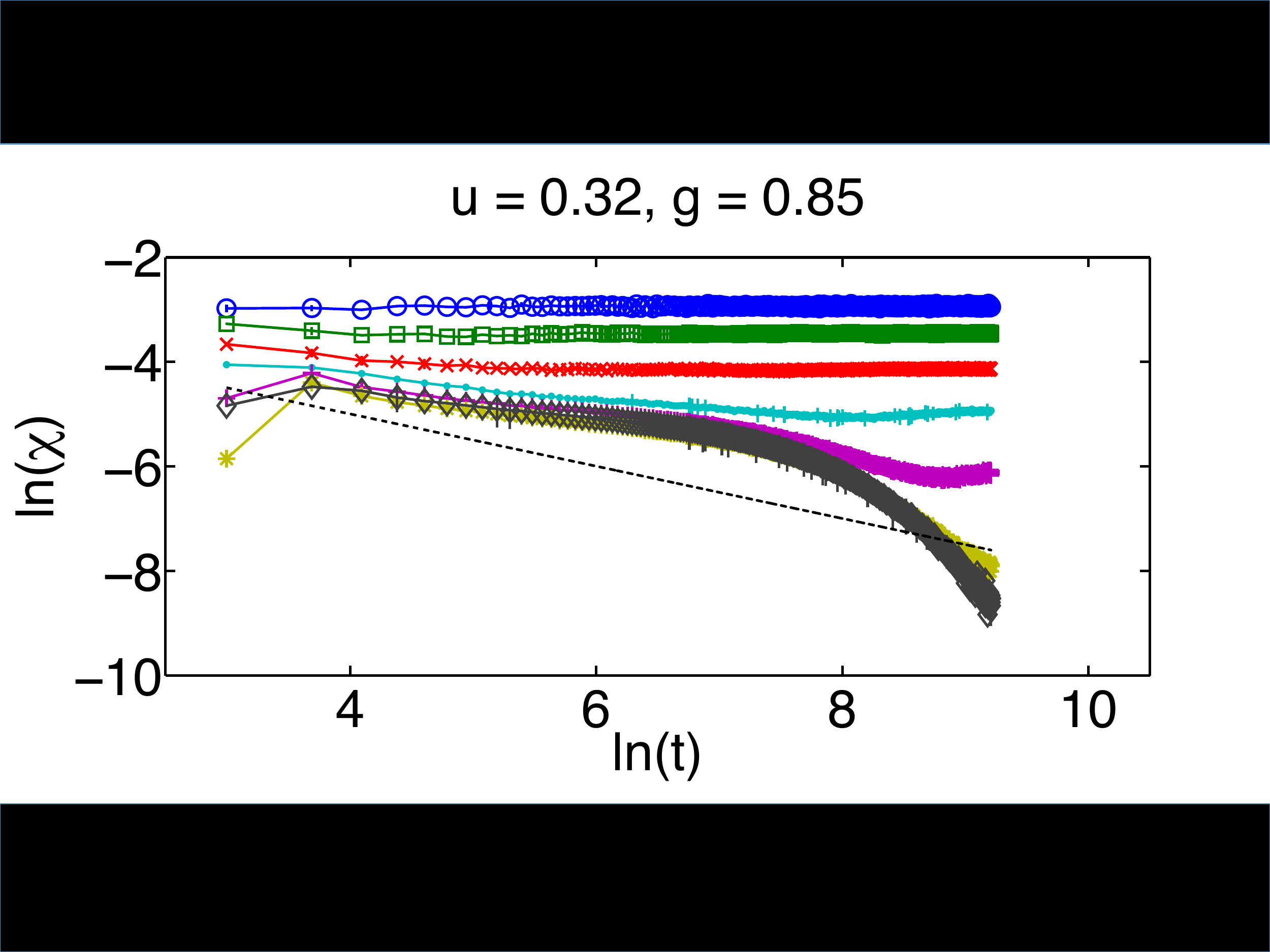}
\end{minipage}\\

\caption{Three characteristic time series for the temporal autocorrelator with $u = 0.32$ and $\Delta t = 1$.  In each panel, we show time series for a particular value of the hopping $g$.  Only a few representative error bars are displayed in each time series.  The legend refers to different lattice sizes $L$.  The reference lines in panels (b) and (c) show diffusive $t^{-\frac{1}{2}}$ decay.}
\label{fig:chits}
\end{figure}

\indent We can discriminate three qualitatively different behaviors of $\chi$ vs.\ $t$ in our interacting model.  Figure \ref{fig:chits} shows examples of each of these behaviors at interaction strength $u = 0.32$.
Panel (a) is characteristic of the low $g$ regime, where the autocorrelator stays invariant over several orders-of-magnitude of time, and there is no statistically significant difference between time series for
different $L$.  At higher $g$, as in panel (b), the time series show approximately power-law decay culminating in saturation to a late-time asymptote.  For the largest systems, the power law is roughly consistent
with the diffusive expectation of $t^{-\frac{1}{2}}$ decay.  The late-time asymptote decays with $L$ (as expected from energy conservation\footnote{The statistical fluctuation of the total energy of the randomly chosen initial configuration is of order $\sqrt{L}$.  Suppose the total energy is conserved by the dynamics. We can write $E/\sqrt{L}=x_0 + h A_0 \cos \theta_0=x_\infty + h A_\infty \cos \theta_\infty$.  Here, the subscripts $0$ and $\infty$ refer to the initial and late-time states, $x_0$ and $x_\infty$ are bounded random numbers capturing the expectation value of interactions (and hopping at late-times), $h$ is the non-random amplitude of the quasiperiodic potential, and $A_0$ and $A_\infty$ are positive bounded amplitudes of the Fourier components at the wavevector $k$ of the quasiperiodic potential.  This ansatz implies a finite correlation between the random phases $\theta_0$ and $\theta_\infty$. Therefore, one of the Fourier modes of $\chi$ remains correlated as $L\to\infty$, and we expect $\chi\sim \frac{1}{L}$ in the ergodic phase.  Note that this argument truly applies only to the energy-conserving parent model.  In fact, in our numerics, there is only partial energy conservation, and energy non-conserving events become more prevalent as $u$, $g$, or $L$ is raised.   This means that $\chi$ will generically decay faster than $\frac{1}{L}$ at large $L$ in the ergodic phase.}
)
suggesting that the power-law decay may continue indefinitely in the thermodynamic limit.
Surprisingly, at still larger $g$, there is a third behavior, exemplified by panel (c).  For the largest lattice sizes, the power-law era is not followed by saturation but by
an extremely rapid decay.  The rapid decay is most evident in the large $g$, large $u$ regime, where the energy density of the parent model (\ref{eq:parentmodel}) is relatively large.  
This implies that this behavior might be tied to the multi-photon processes induced by periodic modulation of the Hamiltonian; correspondingly, it also implies that, for fixed $g$ and $u$, we might be able to 
induce the appearance of the rapid decay by increasing $\Delta t$.  We have tested this numerically, and the results support the connection to the energy non-conserving multi-photon processes.
This suggests that there are only two distinct regimes of the parent model represented in Figure \ref{fig:chits}, differentiated by the $L$ dependence of the asymptotic value of the autocorrelator.  We will proceed 
under this working assumption.

\begin{figure}
\begin{minipage}[b]{0.4cm}
       {\bf (a)}

       \vspace{3.3cm}
\end{minipage}
\begin{minipage}[t]{7.9cm}
       \includegraphics[width=7.8cm]{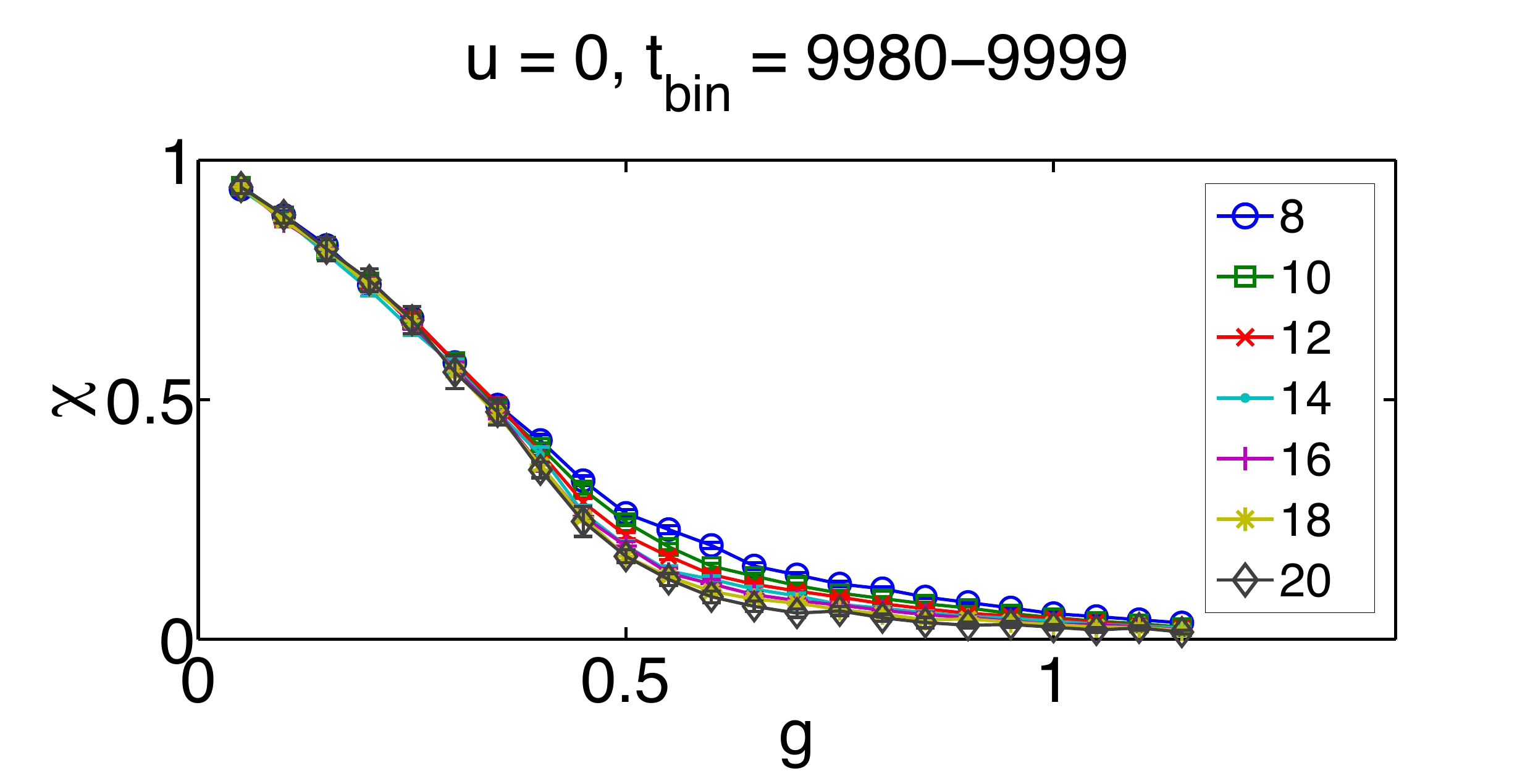}
\end{minipage}\\
\begin{minipage}[b]{0.4cm}
       {\bf (b)}

       \vspace{3.3cm}
\end{minipage}
\begin{minipage}[t]{7.9cm}
       \includegraphics[width=7.8cm]{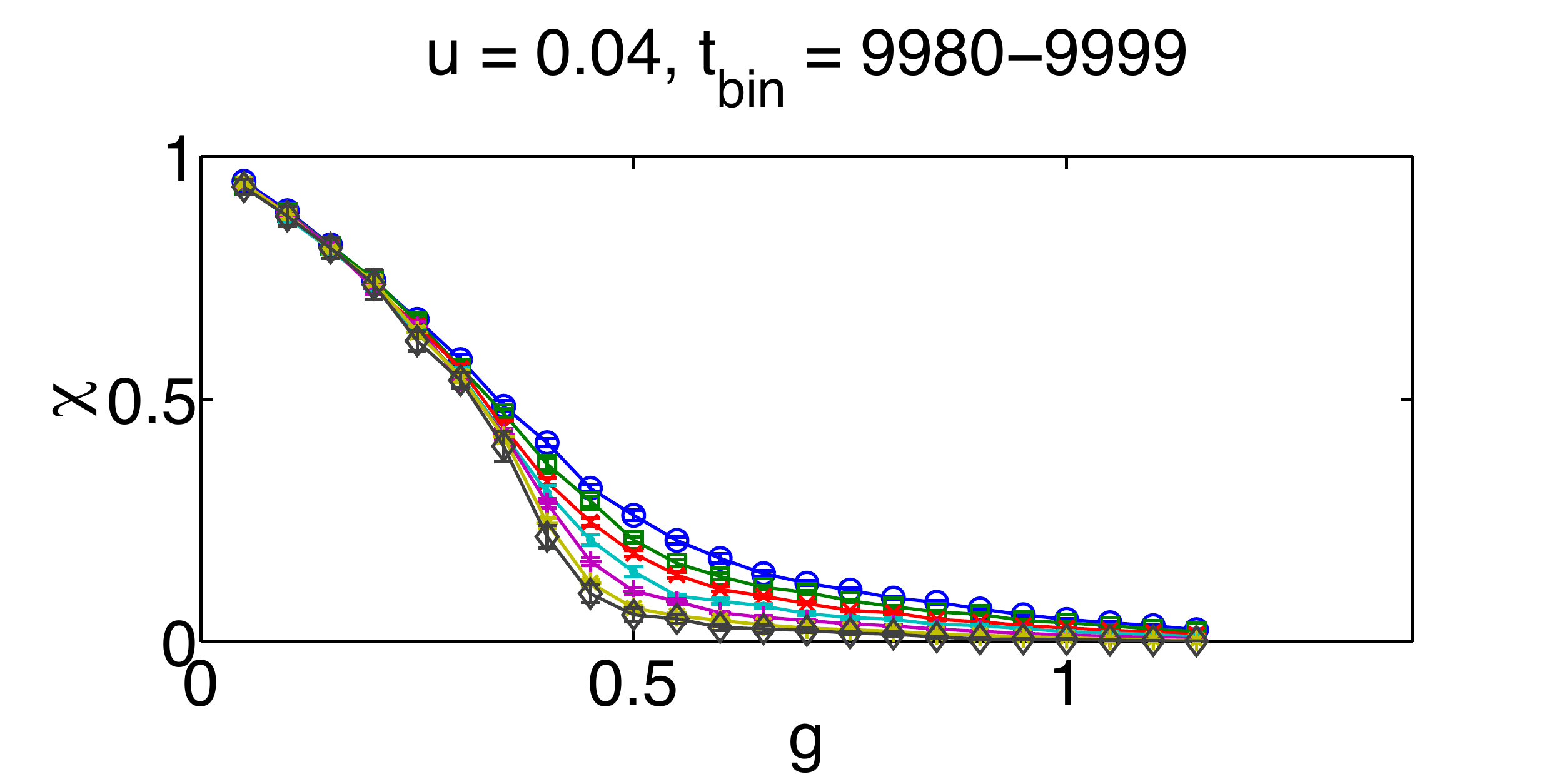}
\end{minipage}\\
\begin{minipage}[b]{0.4cm}
       {\bf (c)}

       \vspace{3.3cm}
\end{minipage}
\begin{minipage}[t]{7.9cm}
       \includegraphics[width=7.8cm]{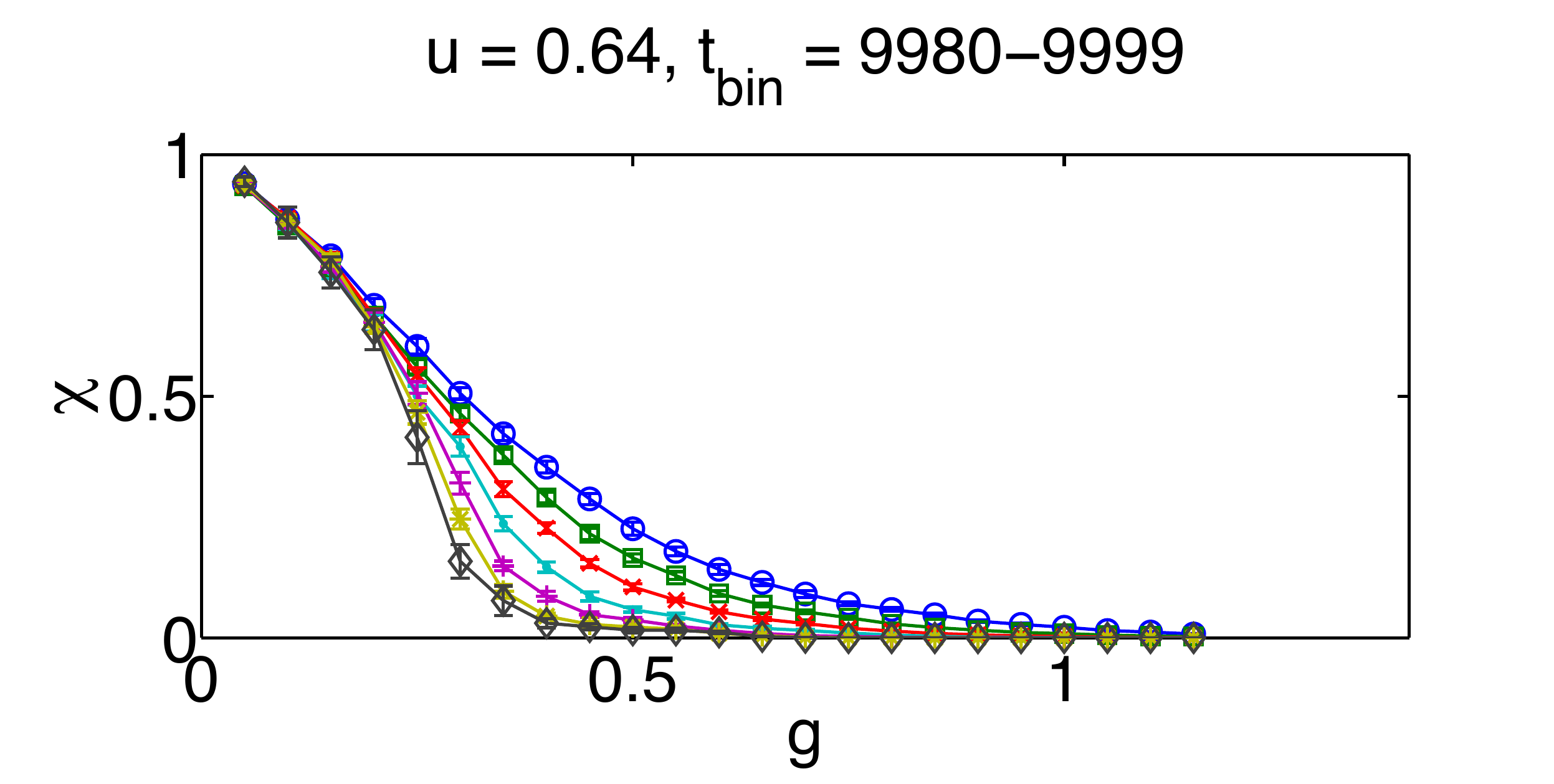}
\end{minipage}\\

\caption{The value of $\chi$ in the latest time bin ($t = 9980\ldots9999$) plotted against $g$.  In panels (a)-(c), $u = 0$, $0.04$, and $0.64$ respectively.  The legend refers to different lattice sizes $L$.}
\label{fig:chicrossings}
\end{figure}

\indent The difference between these two regimes is brought out more clearly in Figure \ref{fig:chicrossings}.  We focus on a late time $t = t_{\text{test}}$ and probe $\chi(t_{\text{test}};L)$ as a
function of $g$ for different lattice sizes.  Panels (a)-(c) show data for $u = 0$, $0.04$, and $0.64$ respectively.  All the panels show a ``splaying" point of the $\chi$ vs.\ $L$ curves, separating
a high $g$ regime in which $\chi(t_{\text{test}};L)$ decays with $L$ from a low $g$ regime in which it does not.
The value of $g$ at this feature decreases monotonically with $u$.  Most importantly, in each case, this value is robust to changing $t_{\text{test}}$; if we halve $t_{\text{test}}$ from the value
that appears in Figure \ref{fig:chicrossings}, the feature appears at approximately the same value of $g$.  
This property of the data is very fortunate: in Section \ref{sec:discussion}.C below, we will use the splaying feature in these plots to put a numerical lower bound on the 
transition value of $g$ for different interaction strengths.  Since time scales get very long near the transition, it is difficult to simulate out to convergence in this regime.  
Nevertheless, the fact that the value of $g$ at the splaying feature remains fixed in time implies that we can deduce the phase structure from our finite-time observations.

\subsection{Normalized participation ratio}

\indent One of the commonly used diagnostics for studying single-particle localization is the inverse participation ratio (IPR).  This quantity is intended to 
probe whether quantum states explore the entire volume of the system and is often defined as the sum over sites of the amplitude of the state 
to the fourth power: $\sum_j |\psi_j|^4$.  Typically, the IPR is inversely proportional to the localization volume $\xi^d$ in a single-particle localized phase and decays to zero as the inverse of the system volume in an extended phase.

\indent We now describe how this quantity can be fruitfully exploited in the many-body context.  Let $c$ denote some specific configuration of $N$ particles in $L$ sites.
Then, we can write the state of the system in the configuration basis as:
\begin{equation}
\label{eq:configurationstate}
\ket{\Psi(t)} = \sum_{\lbrace c \rbrace} \psi_c(t) \ket{c}
\end{equation}
The configuration-basis IPR is simply:
\begin{equation}
\label{eq:participationratio}
P(t;L) \equiv \left[ \sum_c |\psi_c(t)|^4 \right]
\end{equation}
where the square brackets, as usual, denote a sample average.  Interpreting $P(t;L)$ as the inverse of the number of configurations on which $\ket{\Psi(t)}$ has support, we now define
the normalized participation ratio (NPR):
\begin{equation}
\label{eq:participationratioeta}
\eta(t;L) \equiv \frac{1}{P(t;L)V_H}
\end{equation}
The quantity $\eta(t;L)$ then represents the fraction of configuration space that the system explores.
We expect $\eta(t;L)$ to be independent of $L$ at late times in the ergodic phase.  In the many-body localized phase, 
we expect $\eta(t;L)$ to decay exponentially with $L$.

\begin{figure}
\begin{minipage}[b]{0.4cm}
       {\bf (a)}

       \vspace{3.3cm}
\end{minipage}
\begin{minipage}[t]{7.9cm}
       \includegraphics[width=7.8cm]{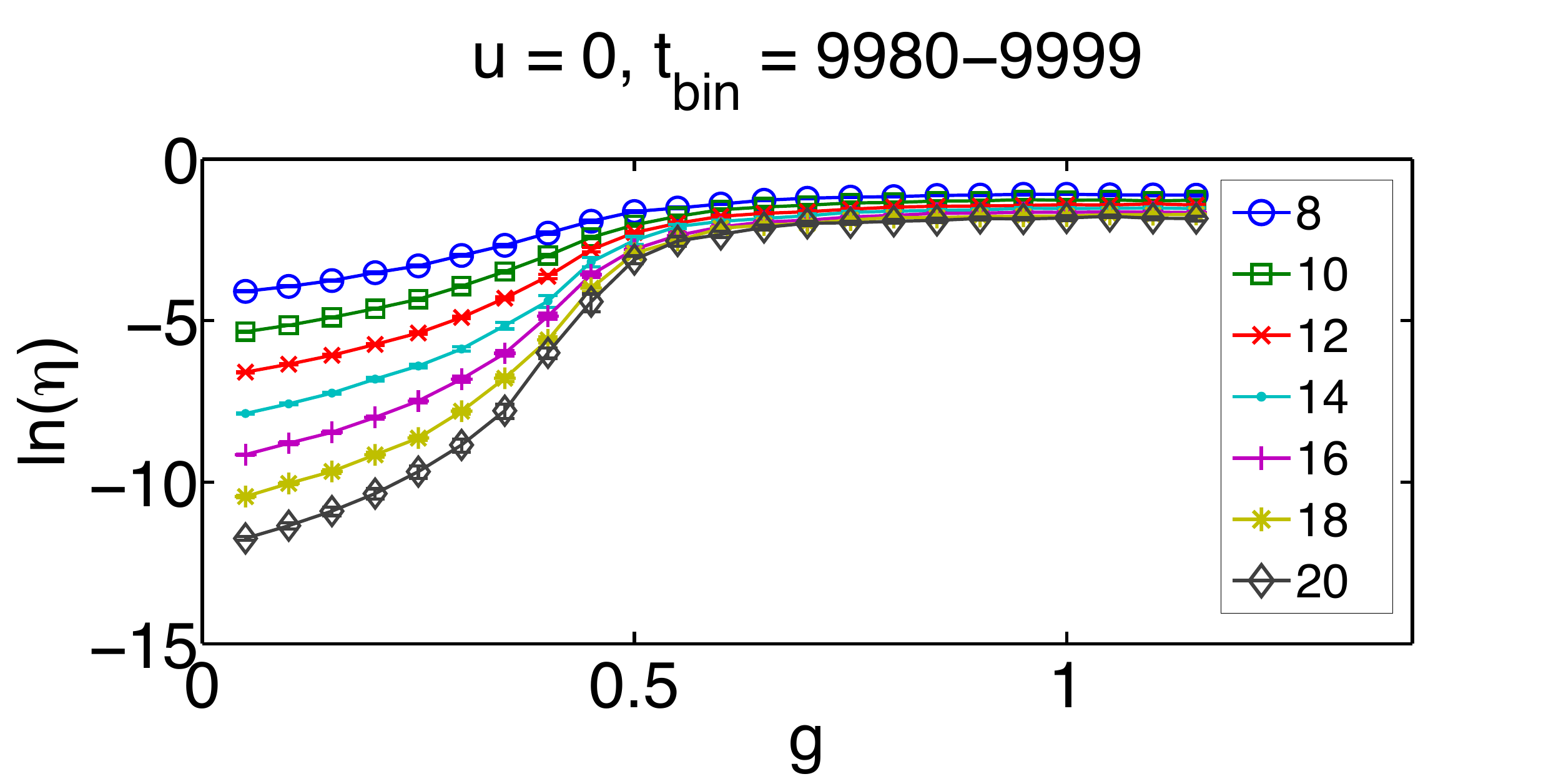}
\end{minipage}\\
\begin{minipage}[b]{0.4cm}
       {\bf (b)}

       \vspace{3.3cm}
\end{minipage}
\begin{minipage}[t]{7.9cm}
       \includegraphics[width=7.8cm]{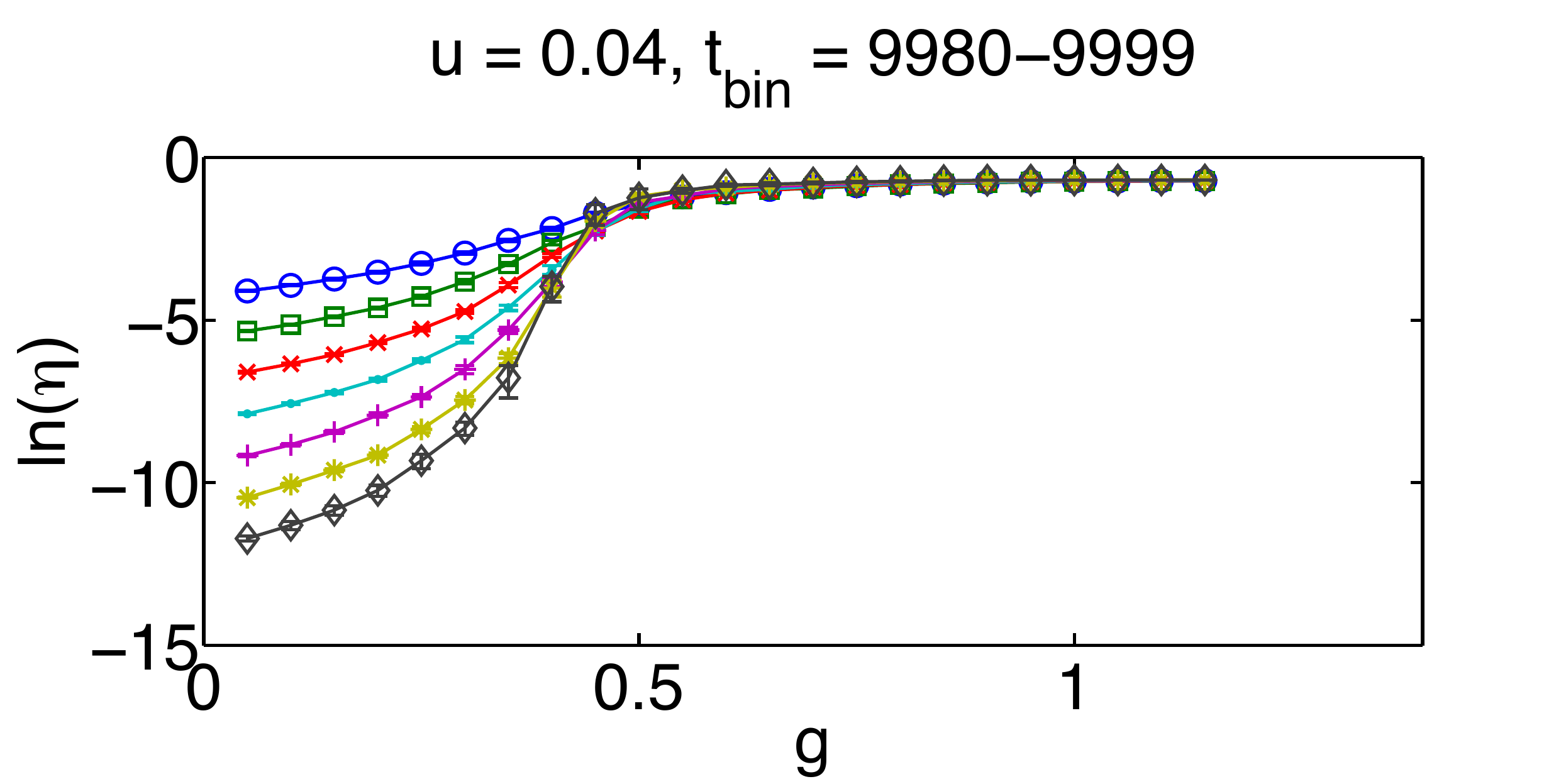}
\end{minipage}\\
\begin{minipage}[b]{0.4cm}
       {\bf (c)}

       \vspace{3.3cm}
\end{minipage}
\begin{minipage}[t]{7.9cm}
       \includegraphics[width=7.8cm]{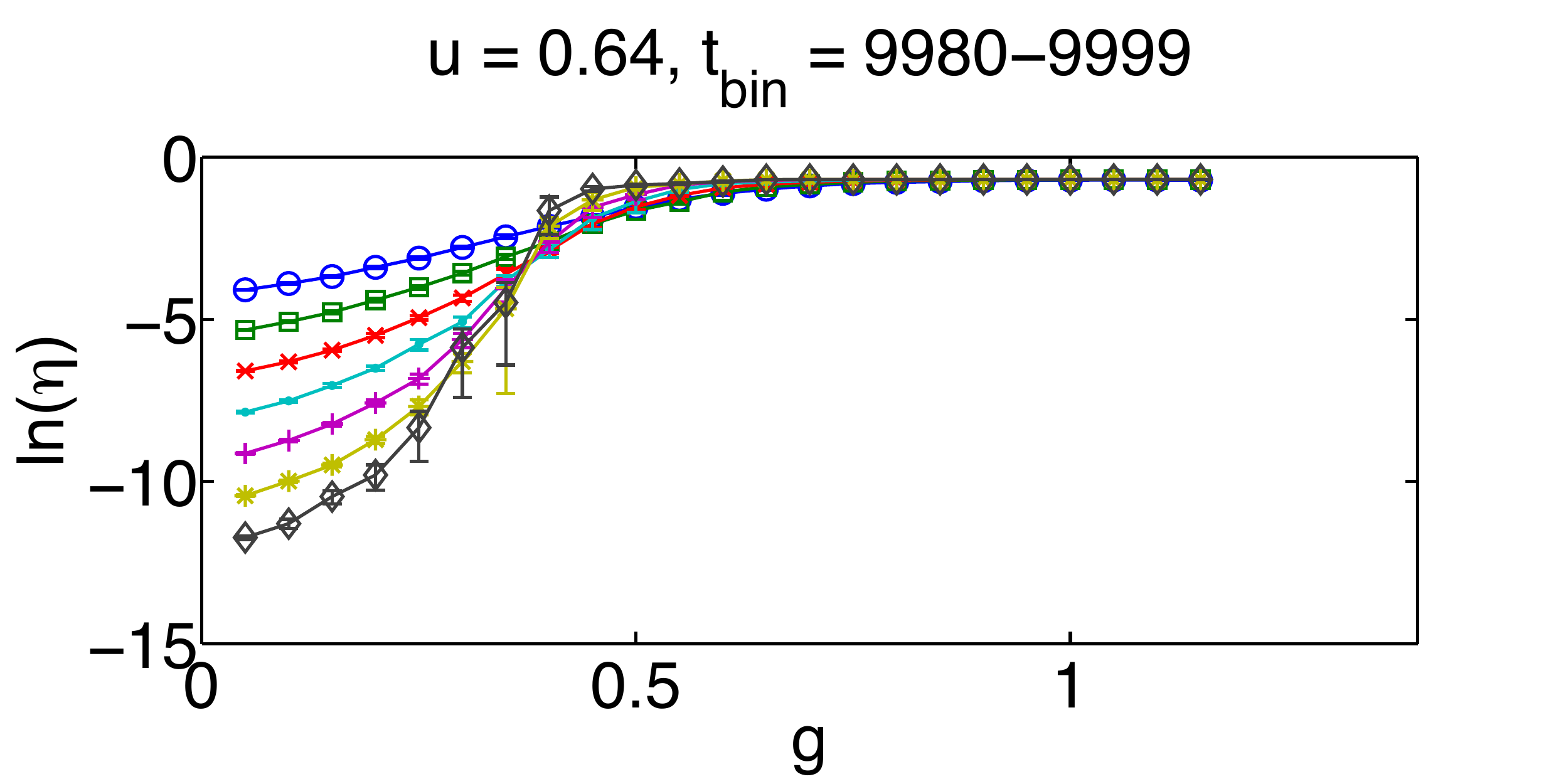}
\end{minipage}\\
\caption{The value of $\eta$ in the latest time bin ($t = 9980\ldots9999$) plotted against $g$.   In panels (a)-(c), $u = 0$, $0.04$, and $0.64$ respectively.  The legend refers to different lattice sizes $L$.  See equation (\ref{eq:participationratioeta}) for the definition of $\eta$. In the ergodic phase $\eta\approx 0.5$.}
\label{fig:etasnaps}
\end{figure}

\begin{figure}
\includegraphics[width=7.8cm]{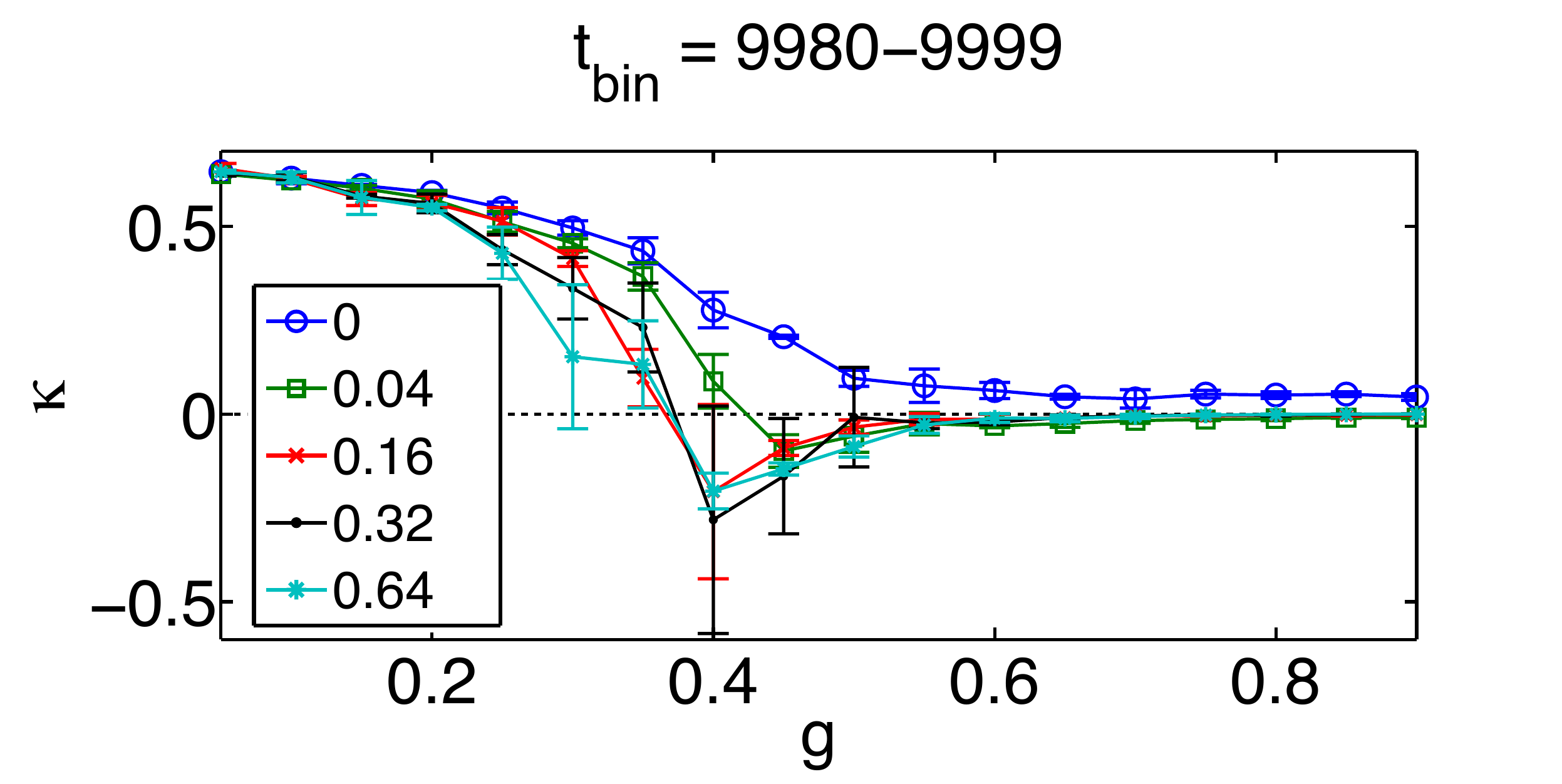}
\caption{Estimates of $\kappa$ from a fit of $\eta \propto e^{-\kappa L}$ in the latest time bin ($t_{\text{bin}} = 9980-9999$).  The legend refers to different values of the interaction strength $u$.}
\label{fig:kappaestimates}
\end{figure}

\indent In Figure \ref{fig:etasnaps}, we plot $\eta(t_{\text{test}};L)$ vs.\ $g$ for $u = 0$, $0.04$, and $0.64$.  The figure reveals an important difference between the non-interacting and interacting models.  At low $g$, both with and without interactions, $\eta$ decays exponentially with $L$:
\begin{equation}
\label{eq:etascaling}
\eta \propto \exp(-\kappa L)
\end{equation}
with $\kappa > 0$.  More surprisingly, $\eta$ also decays with $L$ at \textit{large} $g$ in the non-interacting case; all that happens is that $\kappa$ becomes essentially independent of $g$.  With even small interactions however, $\eta$ becomes system-size independent in the large $g$ regime, following our ansatz for an ergodic phase.  We bring out this point more clearly in Figure \ref{fig:kappaestimates}, in which we extract estimates for the decay coefficient $\kappa$ for various values of the interaction strength.  Thus, the extended phase of the non-interacting AA model appears to be a special, non-ergodic limit.

\indent Before proceeding, we should caution that, in panels (b) and (c) of Figure \ref{fig:etasnaps}, the collapse at high $g$ looks very appealing because of the use of a semilog plot and would not be so striking on a normal scale.  The axes have been chosen to highlight the exponential scaling at low $g$, which would not be as apparent if we simply plotted $\eta$ vs. $g$.  However, regarding the absence of perfect collapse at high $g$, note that the raw data for the IPR differ by several orders-of-magnitude for different values of the lattice size $L$.  Given this, the coincidence of the order-of-magnitude of $\eta$ for different values of $L$ is already a good indication of the proposed scaling, and some corrections to this scaling should be expected given the modest accessible system sizes.

\subsection{R\'{e}nyi Entanglement Entropy}

\indent Unlike the normalized participation ratio, which provides a global characterization of the time-evolved state, bipartite entanglement is arguably a better proxy for whether a part of the system can act as a good heat bath for the rest.  In the many-body ergodic phase, we expect the bipartite entanglement entropy to be a faithful reflection of the thermodynamic entropy.  This implies an extensive entropy, pinned to its thermal infinite temperature value throughout the phase\footnote{This statement should be interpreted with some care.  Quantum entanglement entropy measures, such as the R\'{e}nyi entropy that we define in equation (\ref{eq:Renyientropy}), carry information about the off-diagonal elements in the reduced density matrix.  These terms have no classical analogue and would not be considered in a thermodynamic calculation.  This difference can result in discrepancies in the subleading behavior.  For instance, consider our calculation of the bipartite R\'{e}nyi entropy of the model state $\ket{\Phi}$ in Section \ref{sec:discussion}.A: the quantum R\'{e}nyi entropy is one bit lower than the R\'{e}nyi entropy calculated by classical counting of configurations.  A more precise analogue of the classical entropy would thus be a ``diagonal" entropy in which all off-diagonal elements of the reduced density matrix were neglected.\label{fn:re}}. In contrast, in the many-body localized phase, we expect an extensive but subthermal entanglement entropy.  This expectation is consistent with the results of three recent papers that focus on the behavior of entanglement measures in the many-body localized phase of the disordered problem~\cite{znidaric2008many,bardarson2012unbounded,vosk2012many} .  These papers also study the time dependence of the entropy beginning from an unentangled product state.  In the many-body localized phase, this growth is found to be slow, generically logarithmic in time.  Since our model lacks disorder altogether, it may be interesting to explore the entanglement dynamics here as well.  In what follows, we comment on the dynamics, but we primarily use the late-time entanglement entropy as yet another tool to help distinguish between the many-body localized and ergodic phases. 

\indent Let subsystem A refer to lattice sites $0,1,\ldots\frac{L}{2}-1$, and let subsystem B refer to the remaining sites in the chain.  We can compute the reduced density matrix of subsystem A by beginning with the full density matrix $\hat{\rho}(t) = \ket{\Psi(t)}\bra{\Psi(t)}$ and tracing out the degrees of freedom associated with subsystem B:
\begin{equation}
\label{eq:rhoredAB}
\hat{\rho}_A(t) \equiv Tr_{B} \lbrace \hat{\rho}(t) \rbrace
\end{equation}
The sample-averaged order-2 R\'{e}nyi entropy of subsystem A is then given by:
\begin{equation}
\label{eq:Renyientropy}
S_2(t;L) \equiv \left[-\log_2 \left( Tr_{A} \lbrace \hat{\rho}_A(t)^2 \rbrace \right)\right]
\end{equation}
Both $S_2$ and the standard von Neumann entropy are expected to attain the same values in the ergodic phase; we choose to focus on the former to save on the computational cost of diagonalizing the reduced density matrix (\ref{eq:rhoredAB}).

\begin{figure}
\begin{minipage}[b]{0.4cm}
       {\bf (a)}

       \vspace{3.3cm}
\end{minipage}
\begin{minipage}[t]{7.9cm}
       \includegraphics[width=7.8cm]{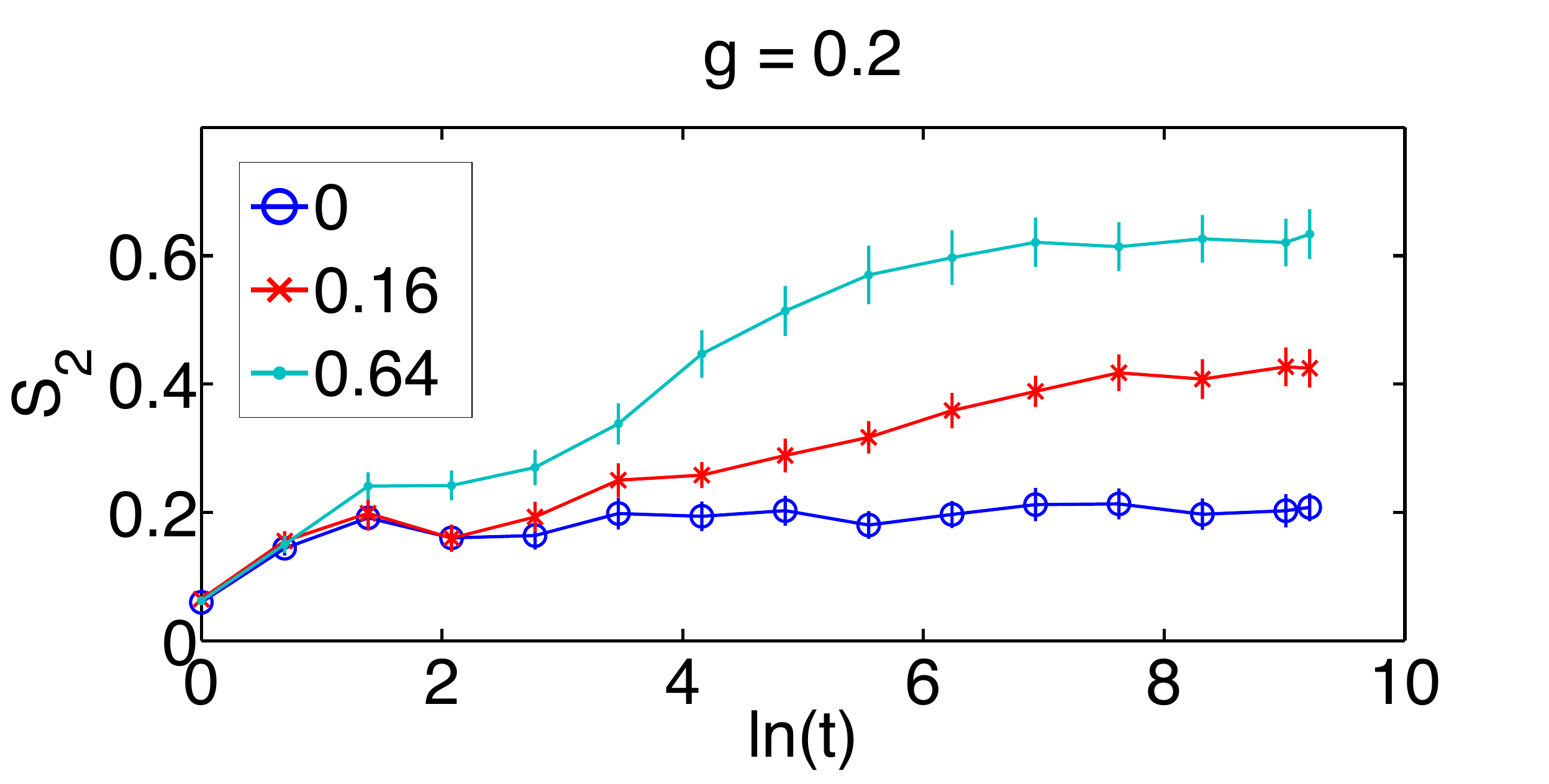}
\end{minipage}\\
\begin{minipage}[b]{0.4cm}
       {\bf (b)}

       \vspace{3.3cm}
\end{minipage}
\begin{minipage}[t]{7.9cm}
       \includegraphics[width=7.8cm]{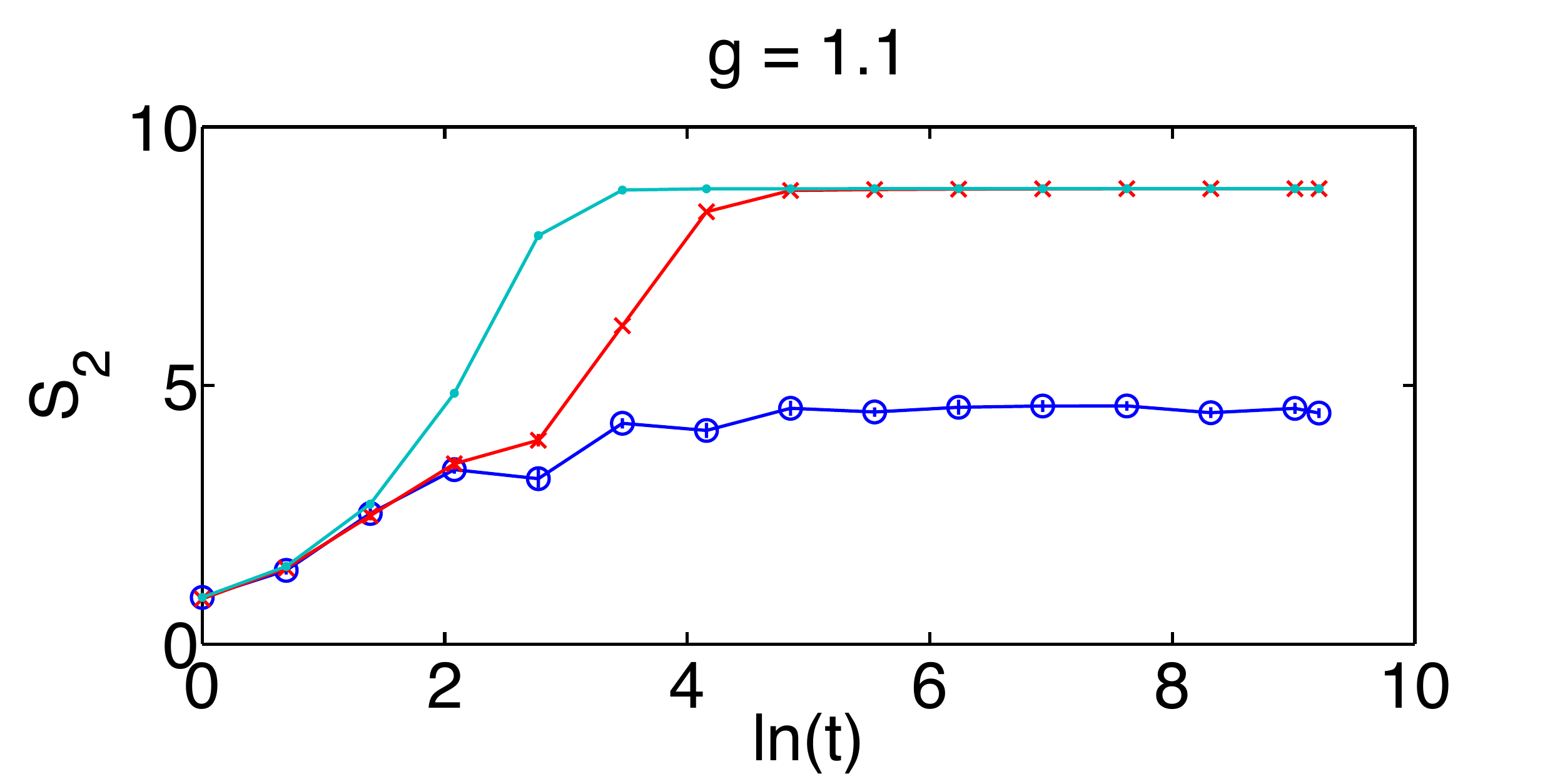}
\end{minipage}\\

\caption{Example time series of the R\'{e}nyi entropy for two values of the tuning parameter $g$.  The legend refers to different values of the interaction strength $u$.  Panel (a) shows data for $L = 10$ lattices at $g = 0.2$.  Panel (b) shows data for $L = 20$ lattices at $g = 1.1$.  In the localized regime, we need to use smaller lattices to see convergence Renyi entropy.}
\label{fig:S2ts}
\end{figure}

\indent Our first task is to examine whether the putative localized phase of our model exhibits the same behavior that was observed with tDMRG
\cite{znidaric2008many,bardarson2012unbounded}.  In panel (a) of Figure \ref{fig:S2ts}, we focus on a low value of $g$ and plot $S_2$ vs. $\ln(t)$ for $L = 10$ lattices.  At very early times, the time series all tend to coincide, reflecting the formation of short-range entanglement at the cut between the subsystems.  Afterwards, the non-interacting time series saturates for several orders-of-magnitude of time, while the interacting time series show behavior that is consistent with logarithmic growth.  In order to clearly establish the saturation that follows the slow growth, we have had to focus on small lattices.
Panel (b) of Figure \ref{fig:S2ts} shows data for large $g$.  Here, the most striking difference between the non-interacting and interacting models lies in the saturation value of the entropy: the interacting model is substantially more entangled, but the saturation value does not appear to depend on the value of $u$.  We will see below that this is another indication that thermalization only occurs in the \textit{interacting}, large $g$ regime.

\begin{figure}
\begin{minipage}[b]{0.4cm}
       {\bf (a)}

       \vspace{3.3cm}
\end{minipage}
\begin{minipage}[t]{7.9cm}
       \includegraphics[width=7.8cm]{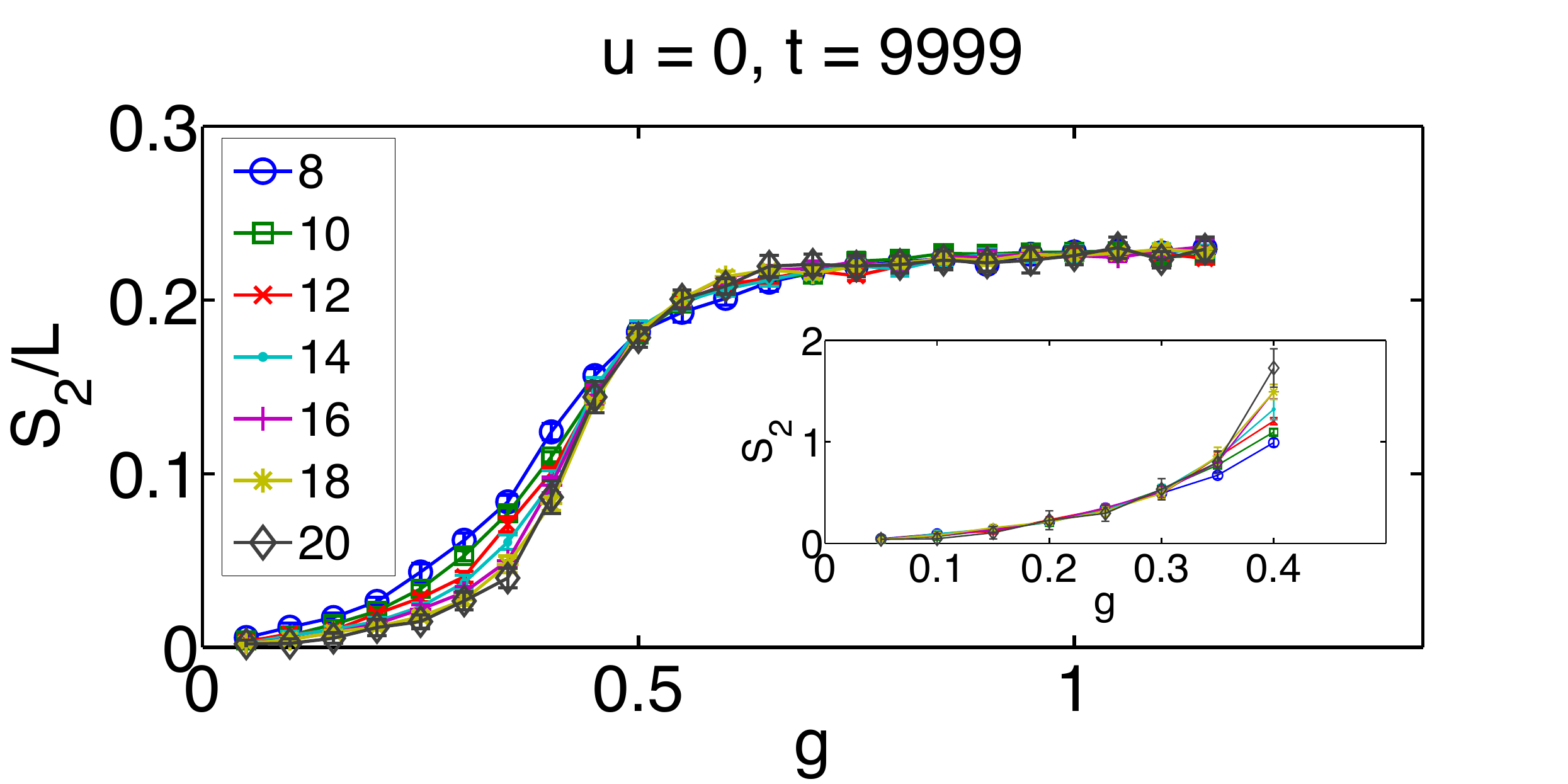}
\end{minipage}\\
\begin{minipage}[b]{0.4cm}
       {\bf (b)}

       \vspace{3.3cm}
\end{minipage}
\begin{minipage}[t]{7.9cm}
       \includegraphics[width=7.8cm]{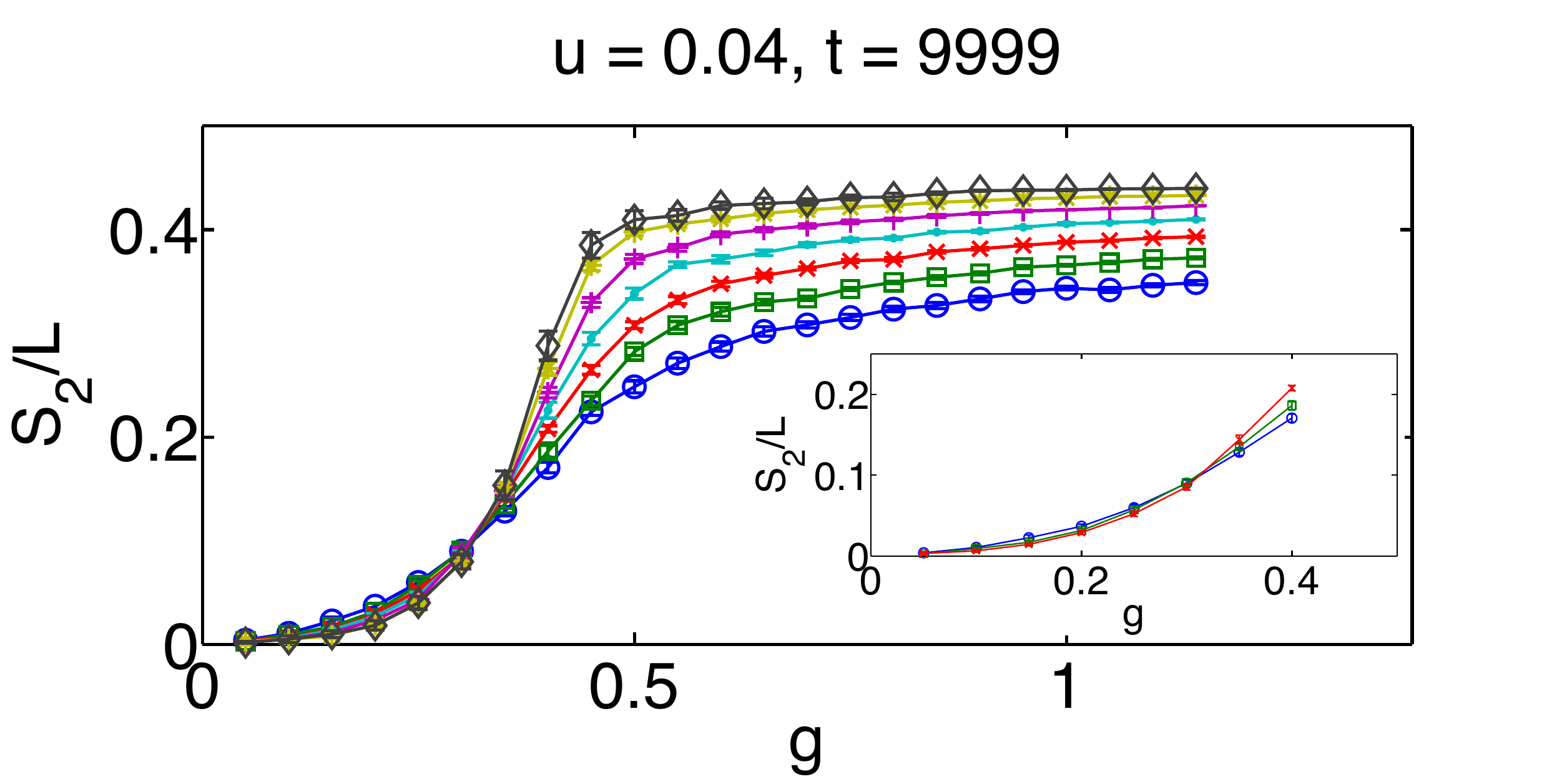}
\end{minipage}\\
\begin{minipage}[b]{0.4cm}
       {\bf (c)}

       \vspace{3.3cm}
\end{minipage}
\begin{minipage}[t]{7.9cm}
       \includegraphics[width=7.8cm]{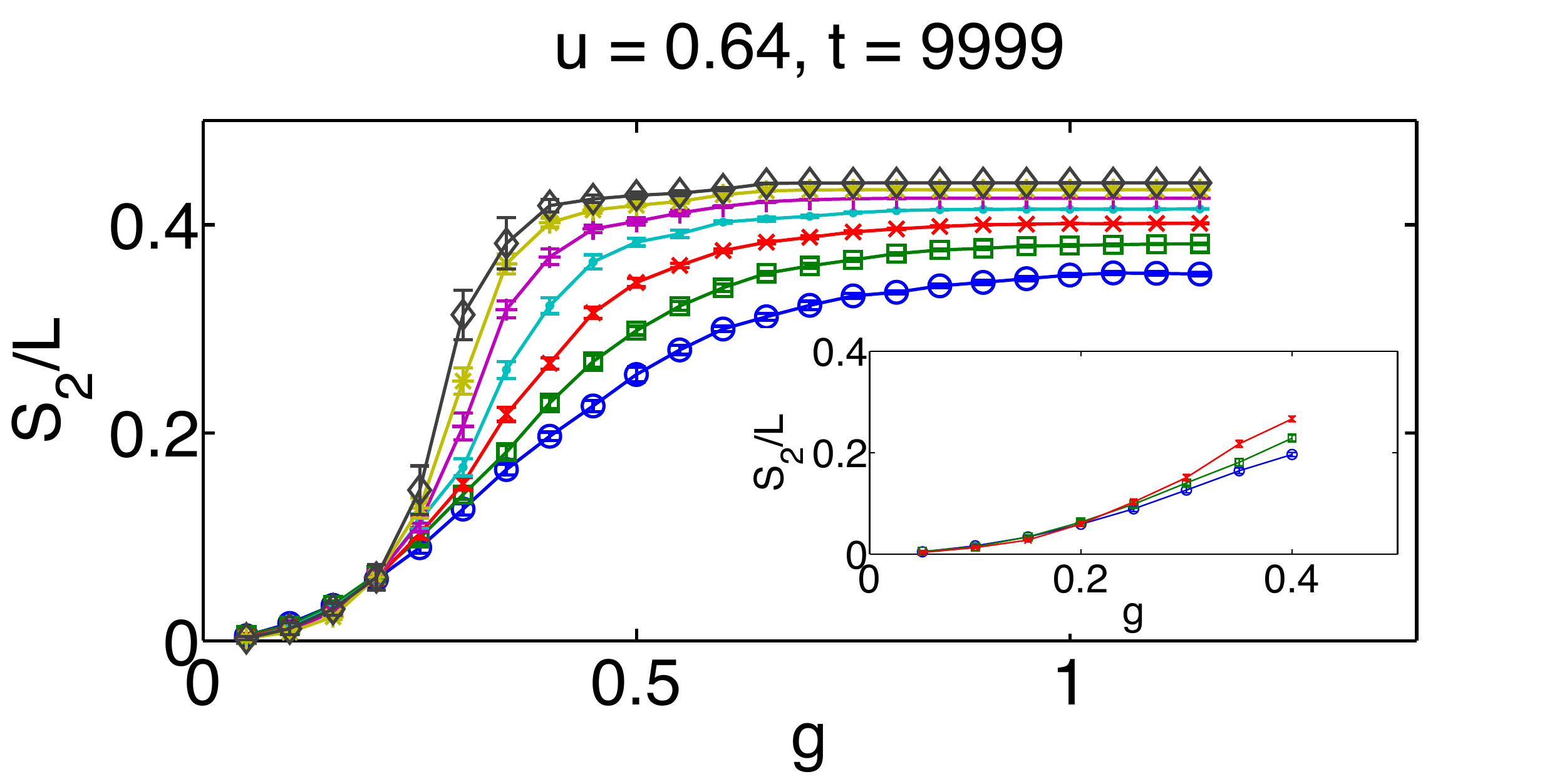}
\end{minipage}\\

\caption{The value of $\frac{S_2}{L}$ at $t = 9999$ plotted against $g$.  In panels (a)-(c), $u = 0$, $0.04$, and $0.64$ respectively.  The legend refers to different lattice sizes $L$.  In panel (a), the inset plot shows $S_2$ vs. $g$ in the low $g$ regime.
In panels (b) and (c), the insets show $\frac{S_2}{L}$ vs. $g$ for low $L$ in the low $g$ regime.}
\label{fig:S2snaps}
\end{figure}

\begin{figure}
\includegraphics[width=7.8cm]{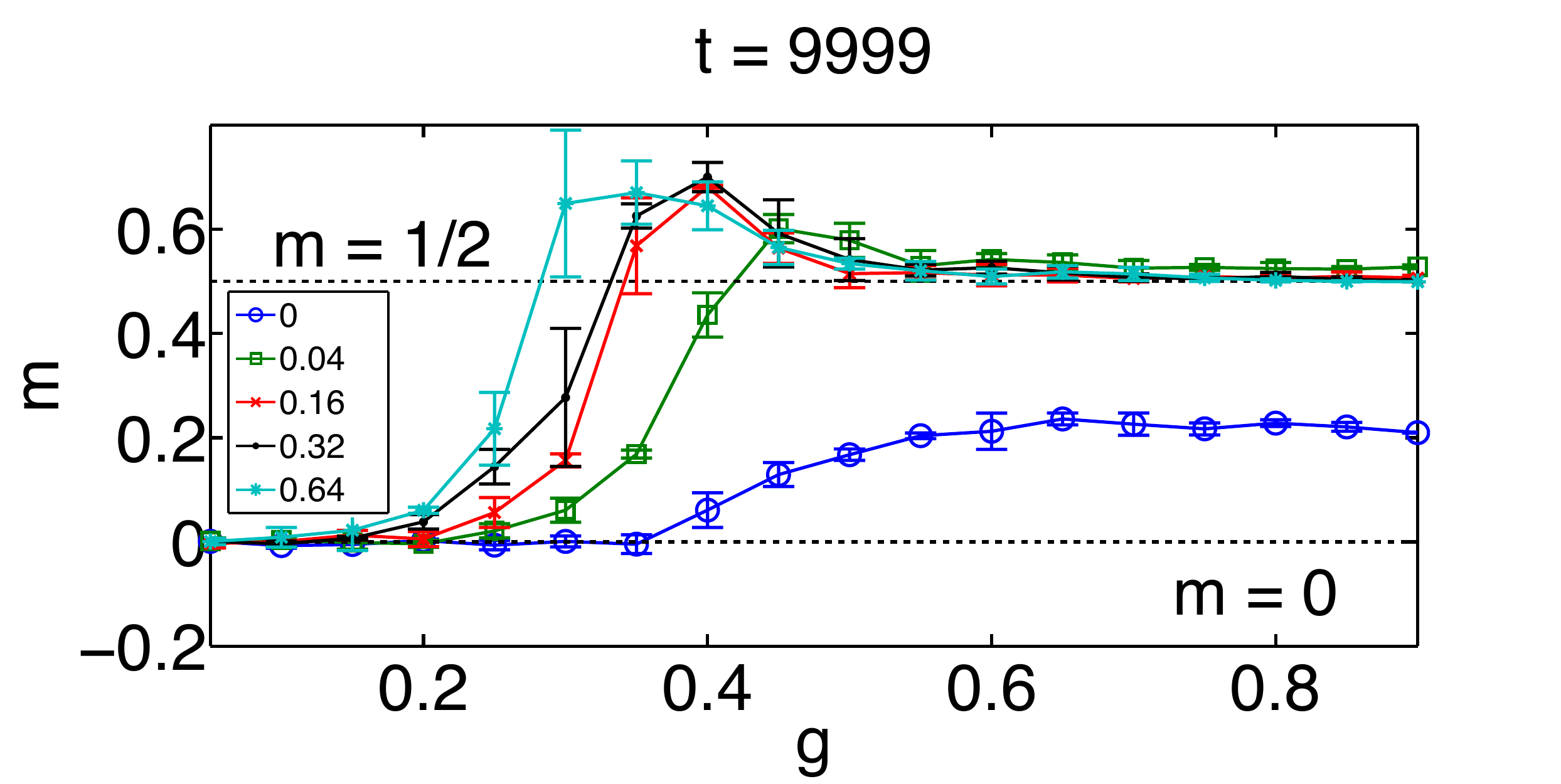}
\caption{The estimated slope of $S_2$ vs. $L$ at late times as a function of $g$.  The legend refers to different values of the interaction strength $u$.}
\label{fig:S2scaling}
\end{figure}

\indent Figure \ref{fig:S2snaps} shows late-time values of the R\'{e}nyi entropy \textit{density} plotted against the tuning parameter $g$.  We first focus on the high $g$ regime.  In panel (a), $u = 0$, and $S_2(t_{\text{test}}; L) \propto L$ for large $g$.  However, the entropy density is less than $\frac{1}{2}$, which is the thermal result when the system has ergodic access to all configurations consistent with particle number conservation.    The situation is dramatically different in panels (b) and (c), where $u = 0.04$ and $0.64$ respectively.  At high $g$, the entropy actually looks \textit{superextensive}.  This is just a finite-size effect, because the entropy is well fit to a linear growth of the form:
\begin{equation}
\label{eq:S2scaling}
S_2(t_{\text{test}};L) = mL-S_{\text{def}}
\end{equation}
where $S_{\text{def}}$ is a constant deficit, typically around $1.15-1.3$.  In Figure \ref{fig:S2scaling}, we show that the slope $m \approx \frac{1}{2}$ at large $g$ in the interacting problem.  This implies that the entropy is thermal in the $L \rightarrow \infty$ limit, where the deficit $S_{\text{def}}$ is negligible.

\indent Now, we turn to the low $g$ regime.  Without interactions, the off-diagonal elements in the reduced density matrix (\ref{eq:rhoredAB}) typically contain only a few frequencies originating from localized single particle orbitals immediately adjacent to the cut.  The number of relevant orbitals is finite in $L$.  As a result, the off-diagonal elements cannot fully vanish, and the reduced density matrix never thermalizes.
The resulting entanglement entropy is independent of $L$ as shown in the inset of panel (a).  In the interacting problem, while the orbitals immediately adjacent to the cut still have roughly the same frequencies, the ``spectral drift" (i.e. the spread of these lines due to sensitivity to the configuration of distant particles) allows for a much larger number of distinct and mutually incoherent contributions to offdiagonal elements of the reduced density matrix.  These off-diagonal elements can dephase more efficiently, leading to a partial thermalization.  This is the mechanism that likely underlies the extensive but subthermal entropy observed by Bardarson et al.\cite{bardarson2012unbounded}.  For small $L$, our numerical results in the low $g$ regime agree well with this expectation.  For larger $L$, the slow dynamics of the entropy formation makes it difficult to observe saturation, both in our work and in the tDMRG study of Bardarson et al.  

\indent If the entropy eventually becomes extensive for all $L$, then the ``crossing" feature that is present in panels (b) and (c) of Figure \ref{fig:S2snaps} would become a ``splaying" feature, with the entropy \textit{density} independent of $L$ at small $g$.  In any case, an interesting property of the data is that the values of $g$ at the crossing features of the $S_2(t_{\text{test}};L)$ vs. $g$ plots are consistent with the locations of the splaying features in the corresponding $\chi(t_{\text{test}};L)$ vs. $g$ plots of Figure \ref{fig:chicrossings}.  This seems to be the case for all $u$.  Thus, these features may be useful in locating the transition.

\section{Modeling the Many-Body Ergodic and Localized Phases}
\label{sec:discussion}

\indent Above, we presented numerical evidence that our interacting AA model contains two regimes that show qualitatively distinct behavior of the autocorrelator, normalized participation ratio, and R\'{e}nyi entropy.  Next, we will propose and characterize model quantum states that qualitatively (and sometimes quantitatively) reproduce the numerically observed late-time behavior in the two regimes.  These model states expose more clearly why the two regimes of our model are appropriately identified as many-body ergodic and localized phases.  

\subsection{The Many-Body Ergodic Phase}

\indent To model the behavior of the putative ergodic phase, we begin by writing down a generic model state in the configuration basis:
\begin{equation}
\label{eq:cstateansatz}
\ket{\Phi} = \sum_{\lbrace c \rbrace} \phi_c \ket{c} = \sum^{\frac{L}{2}}_{n = 0}\sum_{\lbrace c_A,c_B \rbrace} \phi^{(n)}_{AB} \ket{c^{(n)}_A,c^{(n)}_B}
\end{equation}
Here, the $c$ refer to configurations of the full chain, whereas the $c_A$ and $c_B$ refer to configurations of the subsystems A and B, as defined in Section \ref{sec:numerics}.C above.
The superscripts on the configurations and expansion coefficients refer to the number of particles in subsystem A.
Writing the state in terms of the subsystem configurations will be useful shortly, but for now we focus on the statistical properties of the amplitude $\phi_c$.
We assume that this amplitude is distributed as a complex Gaussian random variable:
\begin{equation}
\label{eq:complexGaussian}
p(\phi) = \frac{1}{2\pi \sigma^2} \exp{\left(-\frac{|\phi|^2}{2\sigma^2}\right)}
\end{equation}
Within this distribution, $\langle |\phi|^2 \rangle = 2\sigma^2$ and $\langle |\phi|^4 \rangle = 8\sigma^4$.  From these average values, it is possible to deduce that:
\begin{equation}
\label{eq:cGsigma}
\sigma = \frac{1}{\sqrt{2V_H}}
\end{equation}
for normalization and that the IPR is $P_\Phi = \frac{2}{V_H}$.  This, in turn, implies:
\begin{equation}
\label{eq:cGpr}
\eta_\Phi = \frac{1}{2}
\end{equation}
This result is reproduced \textit{quantitatively} in the numerics in Figure \ref{fig:etasnaps}.

\indent Next, suppose we compute the reduced density matrix of subsystem A in the state $\ket{\Phi}$:
\begin{equation}
\label{eq:ergrhoA}
\hat{\rho}_A = \sum_n \sum_{\lbrace c_A, c_{A'}, c_B \rbrace} \phi^{*(n)}_{AB} \phi^{(n)}_{A'B} \ket{c^{(n)}_A} \bra{c^{(n)}_{A'}}
\end{equation}
To find the R\'{e}nyi entropy, we need to compute the trace of the square of this operator:
\begin{equation}
\label{eq:ergtrrhoA2}
Tr_A{\lbrace \hat{\rho}^2_A \rbrace} = \sum_n \sum_{\lbrace c_A, c_{A'}, c_B, c_{B'} \rbrace} \phi^{*(n)}_{AB} \phi^{(n)}_{A'B} \phi^{*(n)}_{AB'} \phi^{(n)}_{A'B'}
\end{equation}
When we average over our distribution of amplitudes (\ref{eq:complexGaussian}), only the coherent terms survive\footnote{Only the first term on the right-hand side of equation (\ref{eq:ergtrrhoA2_2}) would appear
in a ``classical counting" derivation of the thermodynamic entropy.  The other two terms account for off-diagonal elements in the reduced density matrix (\ref{eq:ergrhoA}).  Please see footnote 53 for more details.}:
\begin{eqnarray}
Tr_A{\lbrace \hat{\rho}^2_A \rbrace} & \approx & \sum_n \sum_{\lbrace c_A, c_B, c_{B'} \rbrace} \langle |\phi^{(n)}_{AB}|^2 |\phi^{(n)}_{AB'}|^2 \rangle \nonumber \\
                                                                 & \quad & + \sum_n \sum_{\lbrace c_A, c_{A'}, c_{B} \rbrace} \langle |\phi^{(n)}_{AB}|^2 |\phi^{(n)}_{A'B}|^2 \rangle \nonumber \\
                                                                 & \quad & - \sum_n \sum_{\lbrace c_A, c_{B} \rbrace} \langle |\phi^{(n)}_{AB}|^4 \rangle
\label{eq:ergtrrhoA2_2}
\end{eqnarray}
The final term accounts for the double counting of terms where $c_A = c_{A'}$ and $c_B = c_{B'}$ simultaneously.  We now introduce the notation:
\begin{equation}
\label{eq:combination}
\gamma(P,Q) = \frac{P!}{Q!(P-Q)!}
\end{equation}
and evaluate the expectation values in equation (\ref{eq:ergtrrhoA2}) to obtain:
\begin{equation}
\label{eq:ergtrrhoA2_3}
Tr_A{\lbrace \hat{\rho}^2_A \rbrace} \approx \frac{2}{V^2_H} \sum_n \gamma \left( \frac{L}{2},n \right)^3
\end{equation}
Finally, using a Stirling approximation to the combination function and a saddle-point approximation for the sum, we find the entropy:
\begin{equation}
\label{eq:ergentropy}
S_{2,\Phi} \approx \frac{L}{2}-\log_2{\left( \frac{4}{\sqrt{3}}\right)} \approx \frac{L}{2} - 1.2
\end{equation}
This is the same form observed in the numerics (\ref{eq:S2scaling}), and the deficit $S_{\text{def}}$ lies in the observed range.
Asymptotically in $L$, the entropy (\ref{eq:ergentropy}) is \textit{maximal}, and this is exactly the expected behavior when the particle number thermalizes.

\indent There is an important caveat to note here: we have argued above that, if multi-photon processes do not completely destroy energy conservation, then this can lead to relic autocorrelations at late times.  This implies that the assumption of independent random amplitudes cannot be exactly correct on a finite lattice.  However, the numerically-observed relic autocorrelations decay with $L$, suggesting that our assumptions get better as the system size grows.  Therefore, in the thermodynamic limit, this phase is truly thermal.

\subsection{The Many-Body Localized Phase}

\indent Our model for the time-evolved state in the localized regime is founded upon the following intuition: there exists a length scale $\xi$, which is analogous to the single-particle localization length and beyond which particles are unlikely to stray from their positions in the initial state.  Then, if we partition our lattice of length $L$ into blocks of size $\xi$, exchange of particles between blocks is less important than rearrangements of the particles within each block.  Consequently, the total number of configurations accessed by the state of the full system is approximately the product of the number of configurations accessed within each block.  This multiplicative assumption should be very safe in a localized phase.  We additionally assume that, within each block, the dynamics completely scramble the particle configuration.  If a certain block of length $\xi$ contains $n$ particles in the initial state, then the time-evolved state contains equal amplitude for each of the possible ways of arranging $n$ particles in those $\xi$ sites.  In keeping with our numerical protocol, we randomly select the initial state from the space of all possible Fock states of a certain global particle number.  Then, a block of $\xi$ sites contains $n$ particles with probability:
\begin{equation}
\label{eq:blocknumber}
w(\xi,n) = \frac{\gamma(\xi,n)}{2^\xi}\left[ 1 + O\left(\frac{\xi^2}{L}\right) \right]
\end{equation}
We will consider the limit $L \gg \xi \gg 1$, where we can approximate the probability by the first term.  The assumptions proposed above motivate writing down a state of the form:
\begin{equation}
\label{eq:locmodel}
\ket{\Lambda} = \frac{1}{\sqrt{M}} \sum\limits_{\lbrace c_1, \ldots c_{\frac{L}{\xi} } \rbrace }^{\sim} z\left(c_1,\ldots c_{\frac{L}{\xi}}\right) \ket{c_1,\ldots c_{\frac{L}{\xi}}}
\end{equation}
where the tilde on the sum indicates that it should only run over configurations that are consistent with the initial distribution of particles among the blocks.
The factors $z$ are complex phases which depend upon the configuration, and $M$ is a normalization which is equal to the total number of configurations represented in the state $\ket{\Lambda}$.

\indent Before beginning our analysis of the state $\ket{\Lambda}$, we should note that, in contrast to our calculations in the ergodic phase, our goal in the localized regime will
be to qualitatively tie the numerically observed large $L$ behavior to the existence of the length scale $\xi$.  Unfortunately, we cannot achieve the quantitative accuracy of the ergodic model state $\ket{\Phi}$ with
the localized toy-model described above.

\indent We begin by estimating the autocorrelator between the initial state and the model time-evolved state $\ket{\Lambda}$.  A non-zero autocorrelator emerges, because each block is only at half-filling \textit{on average}.  Fluctuations away from half-filling (in either direction) yield a positive typical value of the autocorrelator within a block.  Indicating an average over the distribution (\ref{eq:blocknumber}) with an overline, we find the block value $\overline{\chi_{\text{block}}} \approx \frac{1}{L}$.  This is also the average value for the whole system when $L \gg \xi$:
\begin{equation}
\label{eq:locmodelchi}
\chi_\Lambda \approx \frac{1}{\xi}
\end{equation}

\indent Next, to estimate the IPR, we need to compute the normalization factor $M$.  We begin by estimating the number of explored configurations in each block.
The average of the logarithm of the number of explored configurations within a block is:
\begin{equation}
\label{eq:locmodelMblock}
 \overline{\ln(M_{\text{block})}} \approx \ln{\left(\sqrt{\frac{2}{\pi \xi}}2^\xi\right)} - \frac{1}{2}
\end{equation}
Then, using $\overline{\ln{M}} \approx \frac{L}{\xi}\overline{\ln{M_{\text{block}}}}$, we can estimate $M$ itself as:
\begin{equation}
\label{eq:locmodelM}
M \approx e^{\overline{\ln{M}}} \approx 2^L \left( \frac{\pi e \xi}{2} \right)^{-\frac{L}{2\xi}}
\end{equation}
Using this normalization, we can estimate the NPR $\eta_\Lambda$:
\begin{equation}
\label{eq:locmodellambda}
\ln{\eta_\Lambda} \approx -\frac{L}{2\xi}\ln{\left( \frac{\pi e\xi}{2} \right)} + \frac{1}{2} \ln L + \frac{1}{2} \ln \left( \frac{\pi}{2}\right)
\end{equation}
This qualitatively agrees with the numerically observed behavior (\ref{eq:etascaling}) up to subleading corrections, and in the large-$L$ limit:
\begin{equation}
\label{eq:locmodelkappa}
\kappa \approx \frac{1}{2\xi}\ln{\left(\frac{\pi e\xi}{2}\right)}
\end{equation}

\indent Note that equations (\ref{eq:locmodelchi}) and (\ref{eq:locmodelkappa}) imply a relationship between the scaling behaviors of $\chi$ and $\kappa$ in the localized regime.
This relationship is \textit{not} reflected in our numerical data, in part because we cannot truly attain the limit $L \gg \xi \gg 1$.  The numerically computed value of $\kappa$,
for example, can contain finite-size corrections of order $\frac{\ln(L)}{L}$ or $\frac{\xi^2}{L}$.  Also, we must keep in mind that
the state $\ket{\Lambda}$ is just a toy model that does not capture fine details of the time-evolved states in this regime.  Thus, we must be content with reproducing the qualitative
behavior of each measurable quantity individually, without expecting the relationships between these quantities in $\ket{\Lambda}$ to be exactly reproduced in the data.

\indent We now turn to the R\'{e}nyi entropy, the quantity which most strikingly distinguishes between the non-interacting and interacting localized phases.  To examine this quantity, we revert to partitioning the system in half, instead of into blocks of size $\xi$.  As long as $\xi \ll \frac{L}{2}$, the assumptions that we made above about the blocks of size $\xi$ hold even better for the subsystems A and B.  For example, we can assume that there are ``explored sets" of $M_A$ configurations in subsystem A and $M_B$ configurations in subsystem B respectively, with $M = M_AM_B$.  We consider computing the reduced density matrix $\hat{\rho}_A$,  exactly as in equation (\ref{eq:ergrhoA}) above.  If the off-diagonal elements of this density matrix remain perfectly phase-coherent, it can easily be shown that $S^{\text{coh}}_{2,\Lambda} = 0$.   In reality, there will be a local contribution to the entropy from particles straying over the cut between subsystems A and B.  This mimics the situation in non-interacting localized phases.  Alternatively, suppose that dephasing is sufficiently strong that we can proceed by analogy with the ergodic phase, beginning with equation (\ref{eq:ergtrrhoA2}) and keeping only coherent terms as in equation (\ref{eq:ergtrrhoA2_2}).  Thereafter, the calculation for the model localized state $\ket{\Lambda}$ differs from the calculation for $\ket{\Phi}$.  We need to consider the statistics of the configuration probabilities $|\lambda_{AB}|^2$.  For $|\lambda_{AB}|^2 \neq 0$, we need the configurations on both subsystems to lie within their respective explored sets; this occurs in subsystem A, for example, with probability $\frac{M_A}{\gamma(\frac{L}{2},n)}$.  This reasoning leads to the ``dephased" entropy:
\begin{eqnarray}
S^{\text{dp}}_{2,\Lambda} & \approx & - \log_2\left(\frac{1}{M_A}+\frac{1}{M_B}-\frac{1}{M_AM_B}\right) \nonumber \\
                            & \approx & - \log_2\left(\frac{2}{\sqrt{M}}-\frac{1}{M}\right) \nonumber \\
                            & \approx & \frac{1}{2}\left[1-\frac{1}{2\xi}\log_2{\left(\frac{\pi e \xi}{2}\right)}\right]L -1
\label{eq:locentropy}
\end{eqnarray}
where we have additionally made the approximation that typically $M_A \approx M_B \approx \sqrt{M}$.  With only partial loss of coherence, the entropy would lie between these two limiting cases: $S^{\text{coh}}_{2,\Lambda} \leq S_{2,\Lambda} \leq S^{\text{dp}}_{2,\Lambda}$. Thus, dephasing alone, without additional particle transport, can induce an extensive entropy.

\indent Indeed, our numerics support the view that the main difference between the non-interacting and many-body localized phases is the amount of dephasing.  There does not seem to be a qualitative difference in particle transport.  The particle configuration stays trapped near its initial state, even with interactions, and the system does not thermalize.

\section{Tracing the Phase Boundary}
\label{sec:boundary}

\indent in this section, we use the data from Section \ref{sec:numerics} to extract estimates of the phase boundary between the localized and ergodic phases.  Estimating the location of the MBL transition is extremely challenging.  Given the numerically accessible lattice sizes, satisfying finite-size scaling analyses are difficult to perform.  Nevertheless, rough estimates have been made in the disordered problem~\cite{oganesyan2007localization,monthus2010many,pal2010many,de2012many}, so we will now attempt to extract an approximate phase boundary for our model.

\indent We first consider the autocorrelator.  Above, we noted the ``splaying" feature in the late-time plots of the autocorrelator vs. $g$.  The value of $g$ at this feature can be taken as a lower bound for the transition.  For $g$ slightly greater than this value, it is possible that $\chi$ only decays with $L$ because $\xi > L$ for accessible lattice sizes.  Considering two lattice sizes ($L = 16$ and $20$) and finding when their values of $\chi$ deviate, we find the values reported in the first column of Table \ref{tab:gc}.

\indent Next, we consider the fitting parameter $\kappa$ in equation (\ref{eq:etascaling}).  In Figure \ref{fig:kappaestimates}, we see that there is a region where $\kappa < 0$ for finite interaction strength.  Since $\eta \leq 1$, finite-size effects are clearly dominating the estimate in this region.  We can use the value of $g$ where $\kappa$ is minimal to track how this region moves as $u$ is varied.  This yields the second column of the table.

\indent Finally, a similar approach can be applied to extract estimates of $g_c$ from the fits (\ref{eq:S2scaling}).  There exists a region where $m > \frac{1}{2}$, but this is mathematically inconsistent in the thermodynamic limit.  Therefore, if we find the value of $g$ that maximizes $m$, we can again estimate the location of the region dominated by finite-size effects, yielding the final column of Table \ref{tab:gc}.

\begin{table}
\centering
\begin{tabular}{|r||r|r|r|}
\hline
u & $\chi$ & $\kappa$ & $m$ \\
\hline
$0.04$ & $0.35$ & $0.45$ & $0.45$ \\
\hline
$0.16$ & $0.30$ & $0.40$ & $0.40$ \\
\hline
$0.32$ & $0.25$ & $0.40$ & $0.40$ \\
\hline
$0.64$ & $0.25$ & $0.40$ & $0.35$ \\
\hline
\end{tabular}
\caption{Bounds or estimates of the transition value of $g_c$ at various values of $u$ and based on various measured quantities.  The column titled $\chi$ gives a lower bound on the transition value of $g$ based on the autocorrellator.  The remaining two columns give estimates of $g_c$ based on $\kappa$ and $m$, as defined in Sections \ref{sec:numerics}.B and \ref{sec:numerics}.C respectively.  See Section \ref{sec:boundary} for the reasoning behind the estimates.  All values carry implicit error bars of $\pm 0.05$ as that is the discretization of our simulated values of $g$.  This error bar should be interpreted, for instance, as the error on our estimate of the location of the maximum value of $m$.  The error on our estimate of $g_c$ is, of course, much larger.}
\label{tab:gc}
\end{table}

\indent The estimates of the transition value $g_c$ in Table \ref{tab:gc} were obtained using data for the latest time that we simulated (the time bin $t_{\text{bin}} = 9980 \ldots 9999$ for $\chi$ and $\kappa$ and $t = 9999$ for $m$).  However, we have also estimated $g_c$ for data obtained at a half and a quarter of this integration time, finding consistent results.  Thus, the general phase structure of the model is invariant to changing the observation time, even though not all measurable quantities have converged to their asymptotic values.  Consolidating the information from the estimates in Table \ref{tab:gc}, we propose that the phase diagram qualitatively resembles Figure \ref{fig:phasediagram}.

\indent Before proceeding, it is worth noting that our rough estimates of the phase boundary do not make assumptions regarding the character of the MBL transition (i.e. whether it is continuous or first order). In fact,
some of our plots (e.g. panel (c) of Figure \ref{fig:S2snaps}) hint at the possibility of a discontinuous change in $S_2$ as a function of $g$ in the thermodynamic limit.  We are not aware of any results that rule out a first-order MBL transition, so we must keep this possibility in mind.

\section{Conclusion}
\label{sec:conclusion}

\indent Recently, evidence has accumulated that Anderson localization can survive the introduction of sufficiently weak interparticle interactions, giving rise to a many-body localization transition in disordered systems~\cite{basko2006metal,basko2007problem,oganesyan2007localization,pal2010many,de2012many}.  The MBL transition appears to be a thermalization transition: in the proposed many-body localized phase, the fundamental assumption of statistical mechanics breaks down, and the system fails to serve as its own heat bath~\cite{oganesyan2007localization,pal2010many}.  We have presented numerical evidence that this type of transition can also occur in systems lacking true disorder if they instead exhibit ``pseudodisorder" in the form of a quasiperiodic potential.

\indent From one perspective, this may be an unsurprising claim.  For $g < \frac{1}{2}$ the localized single-particle eigenstates of the quasiperiodic Aubry-Andr\'{e} model have the same qualitative structure as those of the Anderson model, so the effects of introducing interactions ought to be similar.  By this reasoning, perhaps it is even possible to guess the phase structure of an interacting AA model using knowledge of an interacting Anderson model: we simply match lines of the two phase diagrams that correspond to the same non-interacting, single-particle localization length.

\indent However, this perspective misses important effects in all regions of the phase diagram.  Most obviously, the AA model has a transition at $u = 0$, and it is interesting to see how this transition gets modified as it presumably evolves into the MBL transition at finite $u$.  It is also important to remember that quasiperiodic potentials are completely spatially correlated.  This means that the AA model lacks rare-regions (Griffiths) effects, and this may have subtle consequences for the dynamics.  Finally, the AA model contains a phase that is absent in the one-dimensional Anderson model, the $g > \frac{1}{2}$ extended phase, and we have seen above that interactions have a profound effect upon this regime.

\indent Understanding MBL in the quasiperiodic context is especially pertinent given the current experimental situation.  Some experiments that probe localization physics in cold atom systems use quasiperiodic potentials, constructed from the superposition of incommensurate optical lattices, in place of genuine disorder.  The group of Inguscio, in particular, has recently explored particle transport for interacting bosons within this setup~\cite{fallani2007ultracold,lucioni2011observation}.  Meanwhile, the AA model has also been realized in photonic waveguides, and experimentalists have studied the effects of weak interactions on light propagation through these systems.  They have also investigated ``quantum walks" of two interacting photons in disordered waveguides~\cite{lahini2009observation,lahini2010quantum}.  This protocol resembles the one we have implemented numerically, so similar physics may arise.  Finally, we note that Basko et al. have predicted experimental manifestations of MBL in solid-state materials.  In such systems, there is always coupling to a phononic bath, so the MBL transition is expected to become a crossover that nevertheless retains interesting manifestations of the MBL phenomena\cite{basko2007possible}.  Whether there exist quasiperiodic solid-state systems to which the predictions of Basko et al. apply remains to be understood.

\indent Given the current experimental relevance of localization phenomena in quasiperiodic systems, we hope that our study will motivate further attempts to understand these issues.  Unfortunately, our ability to definitively identify and analyze the MBL transition is limited by the modest lattice sizes and evolution times that we can simulate.  Vosk and Altman recently developed a strong-disorder renormalization group for dynamics in the disordered problem\cite{vosk2012many}, but the reliability of such an approach in the quasiperiodic context is unclear.  A time-dependent density matrix renormalization (tDMRG) group study of this problem would be a valuable next step.  Tezuka and Garc\'{i}a-Garc\'{i}a have published tDMRG results on localization in an interacting AA model, but their focus was not on the thermalization questions of many-body localization\cite{tezuka2012testing}.  It would be worthwhile to pose these questions using a methodology that allows access to much larger lattices.  However, even tDMRG may have difficulty capturing the highly-entangled ergodic phase \cite{znidaric2008many,bardarson2012unbounded}, so an effective numerical approach for definitively characterizing the transition remains elusive.

\acknowledgments

We thank E.\ Altman, M.\ Babadi, E.\ Berg, S.-B.\ Chung, K.\ Damle, D.\ Fisher, M.\ Haque, Y.\ Lahini, A.\ Lazarides, M.\ Moeckel, J.\ Moore, A.\ Pal, S.\ Parameswaran, D.\ Pekker, S.\ Raghu, A. Rey and J.\ Simon for helpful discussions.  This research was supported, in part, by a grant of computer time from the City University of New York High Performance Computing Center under NSF Grants CNS-0855217 and CNS-0958379.  S.I. thanks the organizers of the 2010 Boulder School for Condensed Matter and Materials Physics.  S.I. and V.O. thank the organizers of the Carges\`{e} School on Disordered Systems.  S.I. and G.R. acknowledge the hospitality of the Free University of Berlin. V.O. and D.A.H are grateful to KITP (Santa Barbara), where this research was supported in part by the National Science Foundation under Grant No. NSF PHY11-25915. V.O. thanks NSF for support through award DMR-0955714, and also CNRS and Institute Henri Poincar\'{e} (Paris, France) for hospitality.  D.A.H. thanks NSF for support through award DMR-0819860.

\appendix

\section{Exact Diagonalization Results for the Single-Particle and Many-Body Problems}
\label{sec:floquet}

\indent This appendix collects exact diagonalization results that supplement the real-time dynamics study in the main body of the paper.

\subsection{Floquet Analysis of the Modified Dynamics}

\indent The goal of the first part of this appendix is to examine the consequences of the modifications to the quantum dynamics described in Section \ref{sec:model}.B above. We first verify that the AA transition survives by diagonalizing the single-particle AA Hamiltonian (i.e. the Hamiltonian (\ref{eq:parentmodel}) with $u = \frac{V}{h} = 0$) and the single-particle unitary evolution operators (\ref{eq:onetimestep}) for various choices of the time step $\Delta t$.  Subsequently, we employ the same approach to examine how varying $\Delta t$ impacts the quasienergy spectrum of the interacting, many-body model.

\subsubsection{Robustness of the Single-Particle Aubry-Andr\'{e} Transition}

\indent To study the single-particle transition, we focus on the inverse participation ratio:
\begin{equation}
\label{eq:spparticipationratio}
P_{\text{sp}}(g;L) = \left( \sum^{L-1}_{j = 0} |\psi_j|^4 \right)
\end{equation}
Here, $\psi_j$ denotes the amplitude of the wave function at site $j$ of an $L$ site lattice.  We enclose the sum in equation (\ref{eq:spparticipationratio}) in parentheses to indicate important differences in the averaging procedure with respect to the many-body inverse participation ratio (\ref{eq:participationratio}).  In the many-body case, we computed the IPR as a sum over configurations in the quantum state at a particular time in the real-time evolution.  Then, we averaged over samples, where a sample was specified by a choice of the offset phase to the potential (\ref{eq:aapotential}) and an initial configuration.  Throughout this appendix, we instead specify a ``sample" solely by the offset phase $\delta$, and we average over \textit{eigenstates} within each sample before averaging over samples.

\begin{figure}
\begin{minipage}[b]{0.4cm}
       {\bf (a)}

       \vspace{3.3cm}
\end{minipage}
\begin{minipage}[t]{7.9cm}
       \includegraphics[width=7.8cm]{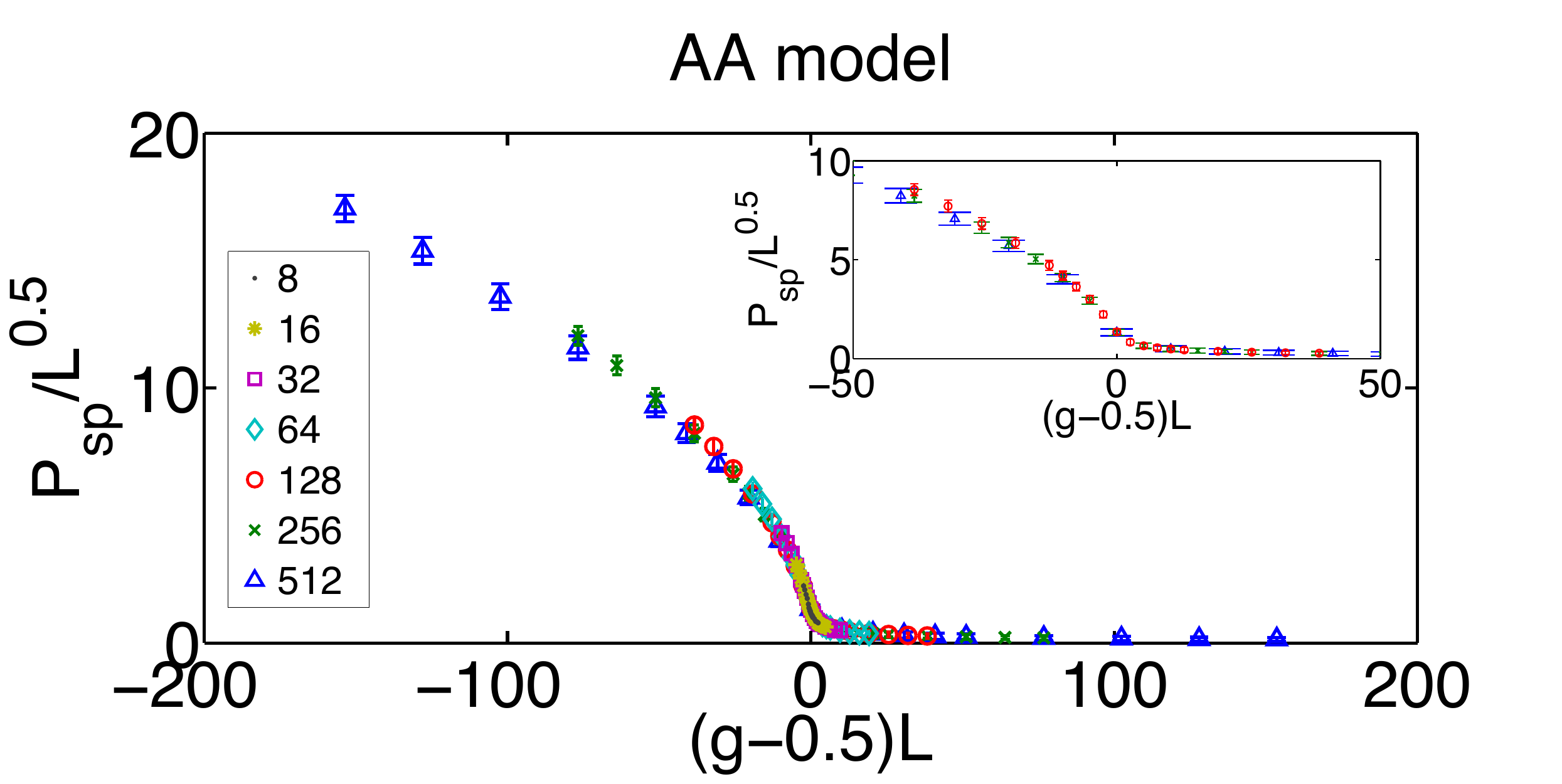}
\end{minipage}\\
\begin{minipage}[b]{0.4cm}
       {\bf (b)}

       \vspace{3.3cm}
\end{minipage}
\begin{minipage}[t]{7.9cm}
       \includegraphics[width=7.8cm]{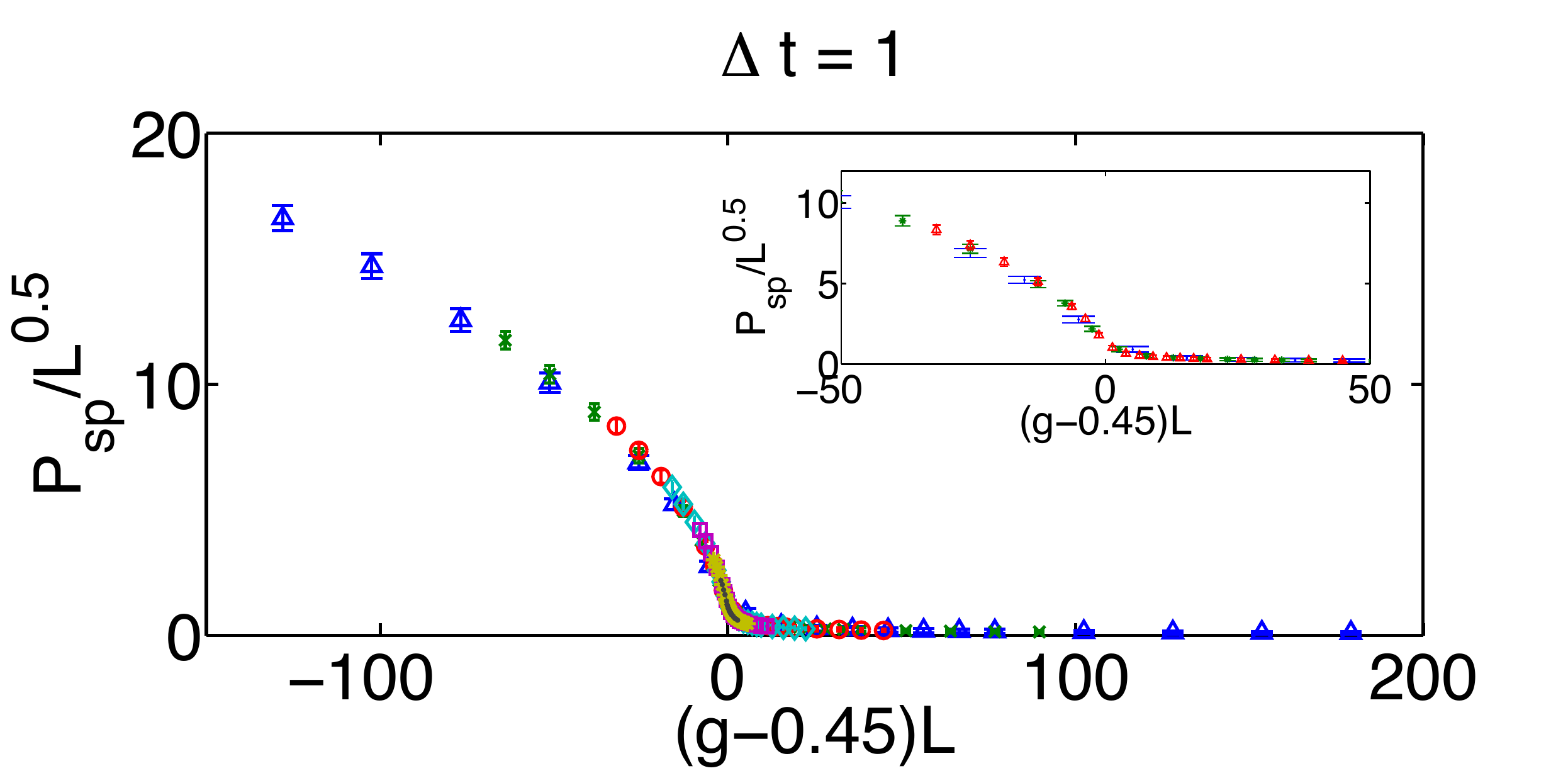}
\end{minipage}\\

\caption{Collapse of single-particle IPR vs.\ g, using the scaling hypothesis (\ref{eq:spprscaling}).  The legend refers to different lattice sizes $L$.  In panel (a), we show data for the usual AA Hamiltonian (\ref{eq:parentmodel}).  In panel (b), we show data obtained from diagonalizing the unitary evolution operator for one time step in the modified dynamics (\ref{eq:onetimestep}).  We use potential wavenumber $k = \frac{1}{\phi}$ and $50$ samples for all lattice sizes.  The insets show magnified views of the curves for the three largest lattice sizes in the vicinity of the transition.}
\label{fig:spcollapse}
\end{figure}

\indent As noted previously, the usual AA model has a transition that must occur, by duality, at $g_c = \frac{1}{2}$.  Near the transition, the localization length is known to diverge with exponent $\nu = 1$\cite{thouless1983wavefunction}.  Our exact diagonalization results indicate that, at the transition, $P_{\text{sp}}(g_c,L) \sim L^{-\frac{1}{2}}$.  Hence, we can make the following scaling hypothesis for the IPR:
\begin{equation}
\label{eq:spprscaling}
P_{\text{sp}} = L^{-\frac{1}{2}} f((g-g_c)L)
\end{equation}
In panel (a) of Figure \ref{fig:spcollapse}, we show that we can use this scaling hypothesis to collapse data for the standard AA model.  We show data for $L = 8$ to $L = 512$, with potential wavenumber $k = \frac{1}{\phi}$ and open boundary conditions.  For all lattice sizes, we average over $50$ samples.

\indent To establish the stability of the AA transition to the modified dynamics, we must ask: can the IPR obtained from diagonalizing the unitary evolution operators (\ref{eq:onetimestep}) be described using the scaling hypothesis (\ref{eq:spprscaling})?   Panel (b) of Figure \ref{fig:spcollapse} shows that this is indeed the case for $\Delta t = 1$.  The only parameter that needs to be changed is $g_c$, which decreases slightly as $\Delta t$ is raised.  This implies that there is a transition in the Floquet spectrum of the system that can be tuned by varying $\Delta t$.  It would be a worthwhile exercise to map out the phase diagram of this single-particle problem in the $(g,\Delta t)$ plane.  We leave this for future work.

\subsubsection{Properties of the Many-Body Quasienergy Spectrum}

\begin{figure}
\begin{minipage}[b]{0.4cm}
       {\bf (a)}

       \vspace{3.3cm}
\end{minipage}
\begin{minipage}[t]{7.9cm}
       \includegraphics[width=7.8cm]{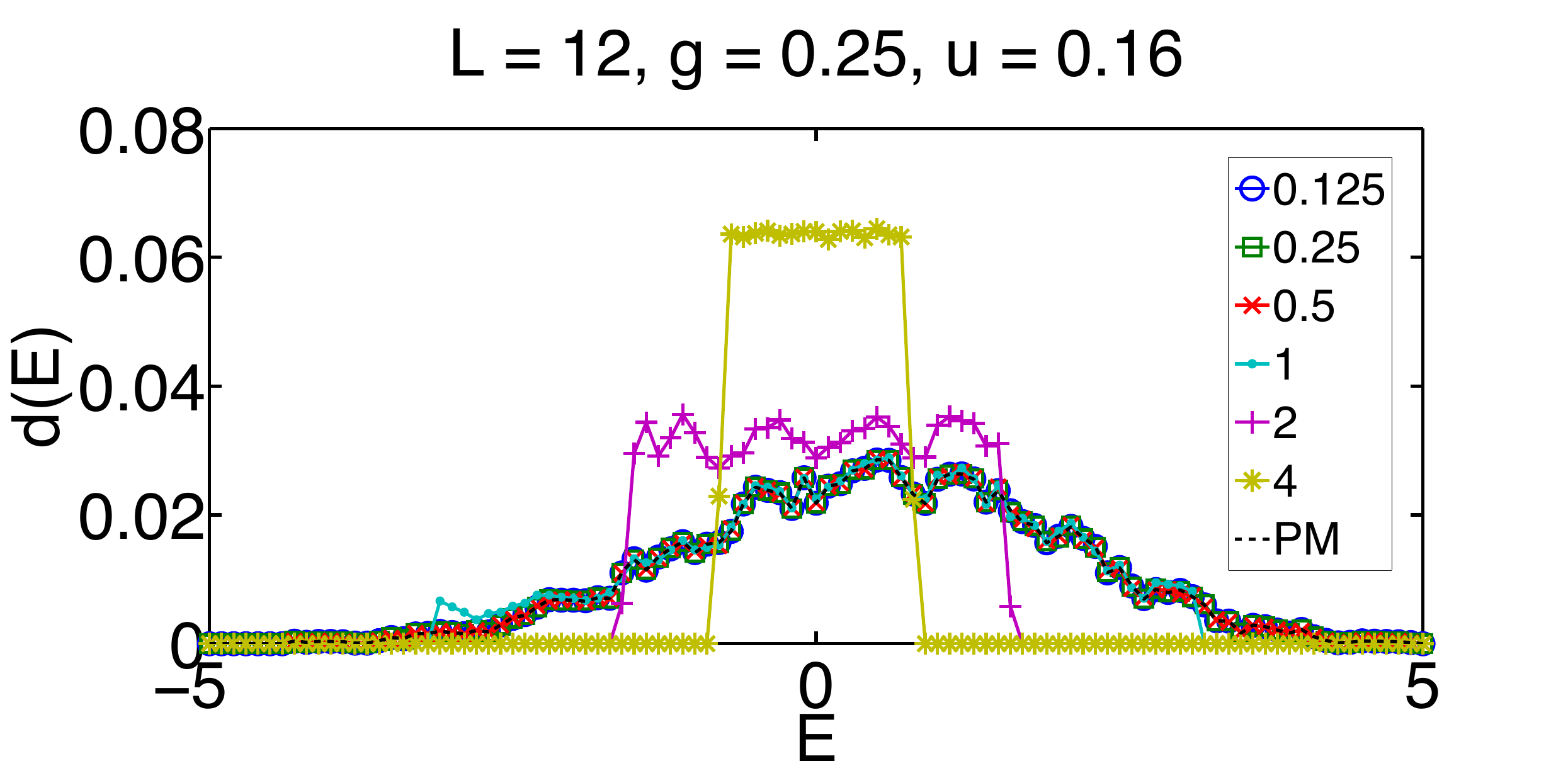}
\end{minipage}\\
\begin{minipage}[b]{0.4cm}
       {\bf (b)}

       \vspace{3.3cm}
\end{minipage}
\begin{minipage}[t]{7.9cm}
       \includegraphics[width=7.8cm]{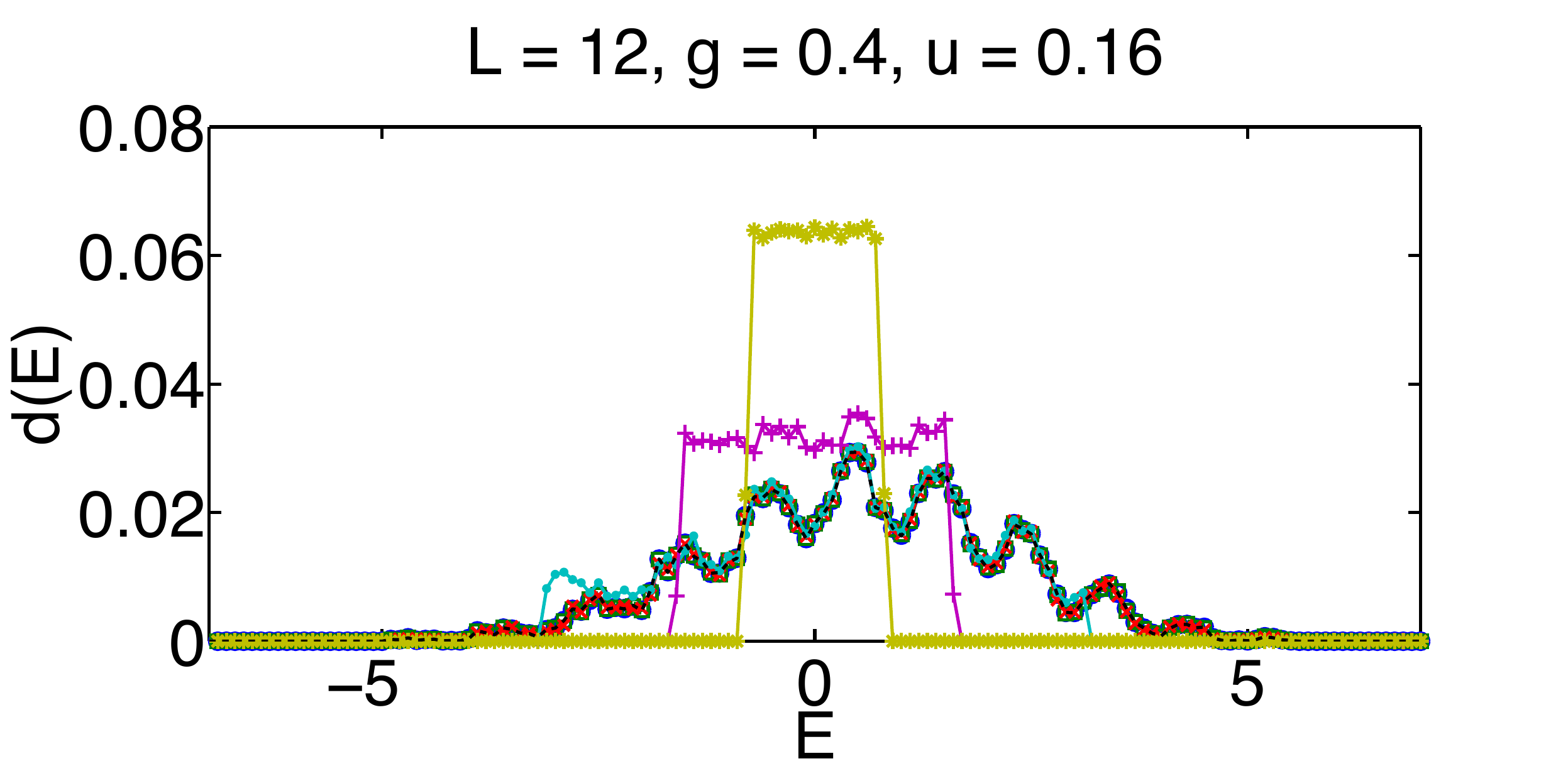}
\end{minipage}\\
\begin{minipage}[b]{0.4cm}
       {\bf (c)}

       \vspace{3.3cm}
\end{minipage}
\begin{minipage}[t]{7.9cm}
       \includegraphics[width=7.8cm]{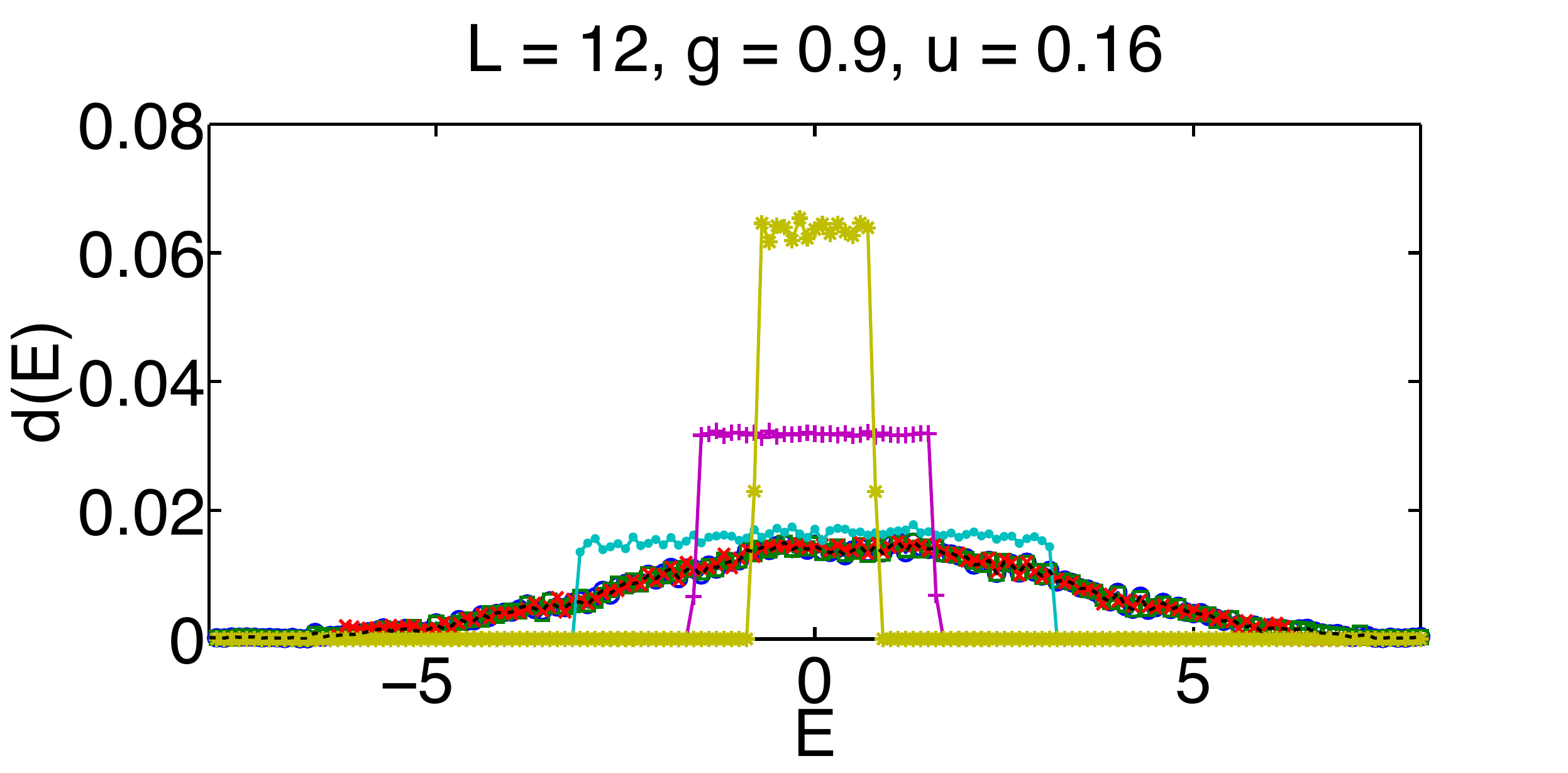}
\end{minipage}\\

\caption{The density-of-states vs.\ quasienergy for $L = 12$ systems at half-filling with interaction strength $u = 0.16$.  The legend refers to different values of $\Delta t$; the time-independent, parent model is referred to as ``PM."  In panels (a)-(c), $g = 0.25$, $0.4$, and $0.9$ respectively.}
\label{fig:quasienergy}
\end{figure}

\indent We now turn our attention back to the effects of the modified dynamics upon the full, many-body model.  In Section \ref{sec:model}.B above, we emphasized that our time-dependent model lacks energy conservation, with multi-photon processes inducing transitions between states of the parent model (\ref{eq:parentmodel}) that differ in energy by $\omega_H = \frac{2\pi}{\Delta t}$.  In this part of the appendix, we will examine how varying $\Delta t$ impacts the quasienergy spectrum of the time-dependent model, using the approach that we applied to the single-particle case above: we diagonalize the time-independent Hamiltonian as well as the unitary evolution operator for one time step of the time-dependent model.  

\indent In Figure \ref{fig:quasienergy}, we plot the density-of-states $d(\Delta t, E)$ in quasienergy space of the parent model and time-dependent models for different values of $\Delta t$.  We focus on $L = 12$ systems at half-filling with fermions (or, since we continue to use the boundary conditions described in Section \ref{sec:model}.C, hardcore bosons).  We fix the interaction strength to $u = 0.16$ and tune $g$ to explore different regimes of the model.  In panels (a)-(c), we plot data for $g = 0.25$, $0.4$, and $0.9$.  According to Table \ref{tab:gc}, these values of $g$ put the system in the localized phase, near the transition, and in the ergodic phase respectively.  

\indent We first consider the consequences of varying $\Delta t$ while holding the other parameters fixed.  For sufficiently small $\Delta t$, the quasienergy spectrum faithfully reproduces all the structure of the energy spectrum of the parent model.  This is unsurprising, because if $\omega_H$ is greater than the bandwidth of the parent model's spectrum, direct multi-photon processes will not take place.  If we now tune $\omega_H$ so that it is less than this bandwidth, the quasienergy spectrum begins to deviate from the parent model's spectrum at its edges.  This effect can be seen, for instance, by examining the trace for $\Delta t = 1$ in panels (a) or (b).   For even higher values of $\Delta t$ (i.e. lower values of $\omega_H$), multi-photon processes strongly mix the states of the parent model, resulting in a flat quasienergy spectrum.

\indent The effect of multi-photon processes can also be enhanced by broadening the parent model's spectrum, which can be achieved by raising $g$ or $u$.  In panel (c) of Figure \ref{fig:quasienergy} for instance, multi-photon processes have significantly flattened the spectrum for $\Delta t = 1$, and deviations from the parent model are even visible for $\Delta t = 0.5$.  Since we always use $\Delta t = 1$ in our real-time dynamics simulations, it is perhaps fortunate that $g = 0.9$ is well within the proposed ergodic phase for $u = 0.16$ and that, near the critical point (i.e. in panel (b)), the quasienergy spectrum for $\Delta t = 1$ still retains much of the structure of the parent model's spectrum.

\indent However, there is one more caveat to keep in mind: the energy content of the system also grows with $L$.  At fixed $g$, $u$, and $\Delta t$, the properties of the parent and time-dependent models deviate from one another as the system size grows.  If we truly want to faithfully reproduce the dynamics of the parent model with the modified dynamics, it may be necessary to scale $\Delta t$ down as we raise $L$.  However, recall that our goal is simply to find MBL in a model qualitatively similar to the parent model (\ref{eq:parentmodel}).  Even with this more modest goal in mind, there is still the danger that, on sufficiently large lattices, multi-photon processes might couple a very large number of localized states and thereby destroy the many-body localized phase of the parent model.  Our numerical observations indicate that this does not happen for the system sizes that we can simulate.  We can keep $\Delta t$ fixed at unity for $L \leq 20$ without issues, accepting the possibility that the sequence of models that we would \textit{in principle} simulate on still larger lattices may require progressively smaller values of $\Delta t$.

\subsection{Level Statistics of the Many-Body Parent Model}

\indent Localization transitions are often characterized by transitions in the level statistics of the energy spectrum\cite{shklovskii1993statistics}.  Two of us previously looked at the level statistics of the disordered problem and identified a crossover from Poisson statistics in the many-body localized phase to Wigner-Dyson statistics in the many-body ergodic phase\cite{oganesyan2007localization}.  The intuition that underlies this crossover is the following: in a localized phase, particle configurations that have similar potential energy are too far apart in configuration space to be efficiently mixed by the kinetic energy term in the Hamiltonian.  Therefore, level repulsion is strongly suppressed, and Poisson statistics hold.  Conversely, in an ergodic phase, there is strong level repulsion which lifts degeneracies, leading to Wigner-Dyson (i.e. random matrix) statistics.

\indent Along the lines of the aforementioned study of the disordered problem, we focus on the gaps between successive eigenstates of the spectrum of the many-body parent model (\ref{eq:parentmodel}):
\begin{equation}
\label{eq:gaps}
\delta_n \equiv E_{n+1}-E_{n}
\end{equation}
and a dimensionless parameter that captures the correlations between successive gaps in the spectrum:
\begin{equation}
\label{eq:rratio}
r_n \equiv \frac{\text{min}(\delta_{n},\delta_{n+1})}{\text{max}(\delta_{n},\delta_{n+1})}
\end{equation}
For a Poisson spectrum, the $r_n$ are distributed as $\frac{2}{(1+r)^2}$ with mean $2\ln(2)-1 \approx 0.386$; meanwhile, when random matrix statistics hold, the mean value of $r$ has been numerically determined to be approximately $0.5295\pm0.0006$\cite{oganesyan2007localization}. 

\begin{figure}
\includegraphics[width=7.8cm]{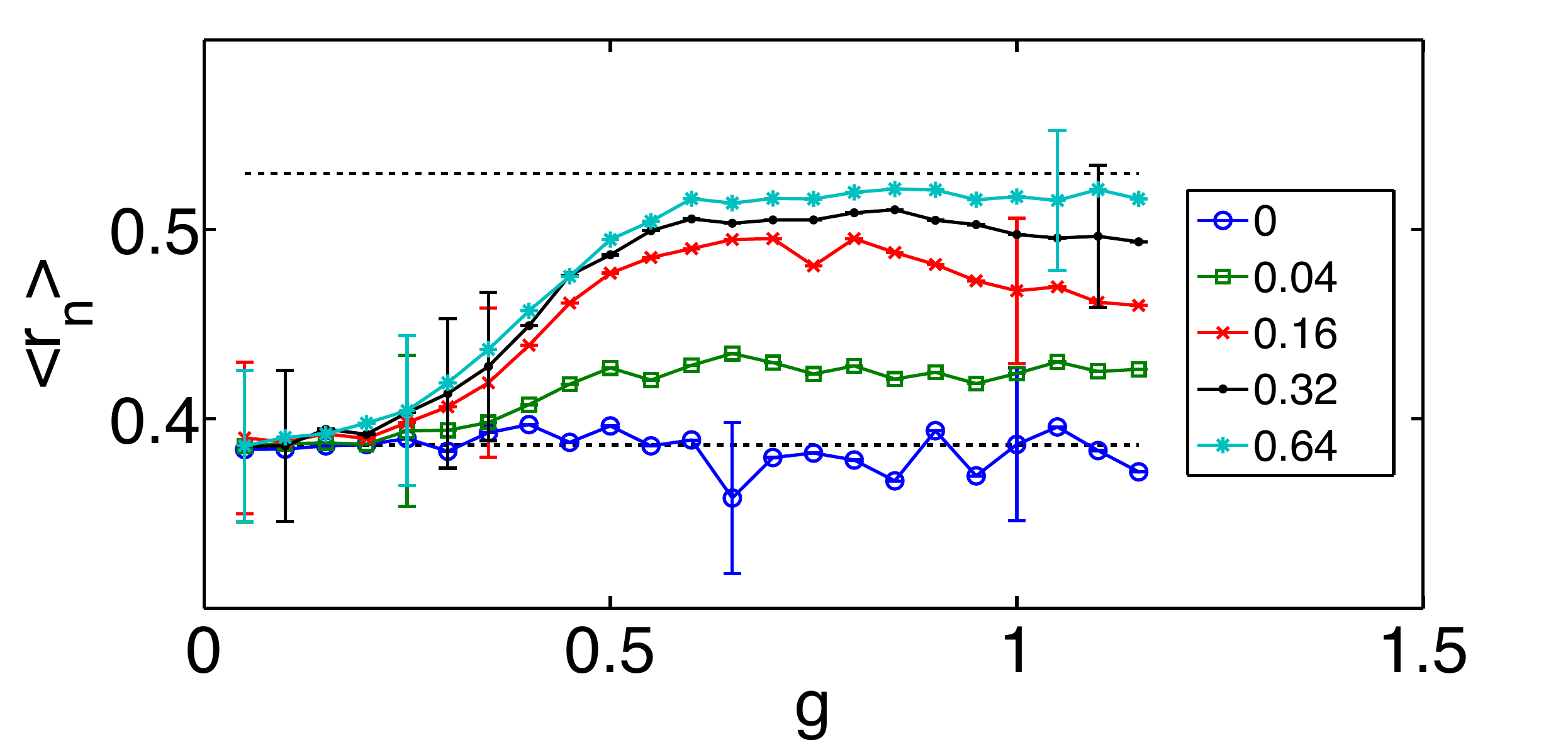}
\caption{The mean of the ratio between adjacent gaps in the spectrum, defined in (\ref{eq:rratio}).  This data was obtained by exact diagonalization of the parent model (\ref{eq:parentmodel}) for $L = 12$ systems.  All data points have been averaged over $50$ samples, and the legend refers to different values of the interaction strength $u$.  The mean value of $\langle r_n \rangle$ shows a crossover from Poisson statistics (indicated by the bottom reference line) to Wigner-Dyson statistics (indicated by the top reference line), for the largest values of $u$.  Representative error bars have been included in the plot; the absent error bars have roughly the same size.}
\label{fig:rstats}
\end{figure}

\indent In Figure \ref{fig:rstats}, we present exact diagonalization results for $L = 12$ lattices at half-filling with potential wavenumber $k = \frac{1}{\phi}$ and the boundary conditions described in Section \ref{sec:model}.C above.  We show data for the same parameter range examined in the body of this paper and average over $50$ samples for each value of $g$ and $u$.  For the largest value of $u$, the mean value of $r_n$ interpolates between the expected values as $g$ is raised, consistent with the existence of a localization transition.  We have also checked that the distributions of $r_n$ have the expected forms in the small and large $g$ limits in this regime.  For smaller values of $u$, we can speculate that $\langle r_n \rangle$ grows with $L$ at large $g$ and approaches the expected value for very large $L$.  To argue for a MBL transition on the basis of exact diagonalization, we would need to study this sharpening of the crossover as $L$ is raised.  This would indeed be an interesting avenue for future work.  For our present purposes however, we only want to check consistency with our real-time dynamics data, as we have done in Figure \ref{fig:rstats}.

\bibliography{draftrefs}

\end{document}